\documentclass[11pt,longbibliography,preprint,prd,nofootinbib,preprintnumbers,amsmath,amssymb]{revtex4}
\pdfoutput=1
\usepackage{dcolumn}% Align table columns on decimal point
\usepackage{bm}% bold math
\usepackage{natbib}
\usepackage{enumerate}
\usepackage{graphicx,wasysym}
\usepackage{hyperref}
\usepackage{comment}

\newcommand{\bea}{\begin{eqnarray}}
\newcommand{\ea}{\end{eqnarray}}
\newcommand{\eea}{\end{eqnarray}}
\newcommand{\beq}{\begin{equation}}
\newcommand{\be}{\begin{equation}}
\newcommand{\ee}{\end{equation}}
\newcommand{\eq}{\end{equation}}
\newcommand{\bi}{\begin{itemize}}
\newcommand{\ei}{\end{itemize}}

\newcommand{\nn}{\nonumber}

\newcommand{\vp}{\varphi}

\newcommand{\p}{\partial}

\newcommand{\half}{\frac{1}{2}}

\newcommand{\el}{ \\}
\newcommand{\bskip}{\;\;\;\;\;\;\;\;\;\;\;}
\newcommand{\Rthree}{$\mathbb{R}^3$ }
\newcommand{\sd}{\dot{\sigma}}
\newcommand{\sdd}{\ddot{\sigma}}
\newcommand{\vpdh}{\dot{\varphi}_H}
\newcommand{\vpddh}{\ddot{\varphi}_H}
\newcommand{\vpdl}{\dot{\varphi}_L}
\newcommand{\vpddl}{\ddot{\varphi}_L}
\newcommand{\vpdddl}{\dddot{\varphi}_L}
\newcommand{\vpl}{{\varphi}_L}
\newcommand{\vph}{{\varphi}_H}

\newcommand{\dt}{\delta \theta}
\newcommand{\cM}{{\cal M}}
\newcommand{\cD}{{\cal D}}

\begin{document}
\preprint{DAMTP-2012-18  \; MIFPA-12-08 \; NSF-KITP-11-258}
\title{Decoupling Survives Inflation: \\ A Critical Look at Effective Field Theory Violations During Inflation}

\author{\large Anastasios Avgoustidis$^{a,e}$}
\author{Sera Cremonini$^{a,b}$}
\author{Anne-Christine Davis$^{a}$}
\author{Raquel H. Ribeiro$^{a}$}
\author{Krzysztof Turzy\'{n}ski$^{c}$}
\author{Scott Watson$^{d}$}
\vspace{1.3cm}

\affiliation{$^{a}$D.A.M.T.P., Cambridge University, Cambridge, CB3 0WA, UK\\
	$^{b}$George and Cynthia Mitchell Institute for Fundamental Physics and Astronomy,
Texas A \& M University, College Station, TX 77843, USA 	\\
$^{c}$Institute of Theoretical Physics, Faculty of Physics, University of Warsaw, ul. Ho\.{z}a 69, 00-681 Warsaw, Poland \\
$^{e}$ School of Physics and Astronomy, University of Nottingham, University Park, Nottingham NG7 2RD, UK\\
$^{d}$Syracuse University, Syracuse, NY 13244, USA}
\date{\today}

\begin{abstract}
We investigate the validity of effective field theory methods and the decoupling of
heavy fields during inflation.
Considering models of inflation in which the inflaton is coupled to a heavy (super-Hubble) degree of freedom
initially in its vacuum state, we find that violations of decoupling are absent unless
there is a breakdown of the slow-roll conditions.  Next we allow for a
temporary departure from inflation resulting in a period of non-adiabaticity during which effective field theory methods
are known to fail.
We find that the locality of the event and energy conservation lead to a tight bound on the
 size of the effects of the heavy field. We discuss the implications for the power spectrum and
 non-gaussianity, and comment on the connection with recent studies of the dynamics of multi-field inflation models.
 Our results further motivate the use of effective field theory methods to characterize cosmic inflation,
and focus the question of observability of additional degrees of freedom during inflation to near the Hubble scale or below --
as anticipated from the Wilsonian notions of decoupling and naturalness.
\end{abstract}

\maketitle
\newpage
%%%%%
\tableofcontents
%%%%%
\newpage

\section{Motivation and Summary of Results}

As the accuracy of cosmological measurements continues to improve at a rapid pace,
it is crucial to establish a robust and economic way to connect data with fundamental theory.
In particle physics and condensed matter systems the framework of Effective Field Theory (EFT)
has proven to be very successful at this endeavor (for reviews see \cite{Kaplan:2005es,Burgess:2007pt}).
However, it has been argued by some that these techniques may fail when implemented in cosmology,
e.g. because of the time evolution of the cosmological background or due to our lack of knowledge of physics
near the Planck-scale (see \cite{Brandenberger:2007qi} for a critical review).
Others have argued that whatever the relevant quantum theory of gravity may be, it should be unitary and causal,
and therefore `trans-Planckian' effects will decouple at energies of observational interest,
in agreement with Wilsonian intuition \cite{Kaloper:2002cs,Kaloper:2002uj}.
Thus far, the second viewpoint seems to be favored since no failures of decoupling have been detected experimentally.

More recently, EFT methods have been further developed for application in both early and late universe
cosmology \cite{Weinberg:2008hq,Creminelli:2008wc,Park:2010cw,Cheung:2007st,Senatore:2010wk,O'Connell:2011ng}.
The approach advocated in \cite{Cheung:2007st,Senatore:2010wk} is particularly powerful, whereby considering the EFT of
perturbations around a given background, the authors capture inflationary models with small sound speed
and therefore the potential for a large, observable level of non-Gaussianity.
This technique has already proven useful for establishing new shapes of non-gaussianity,
and therefore new possibilities for observation \cite{Senatore:2010jy,Senatore:2009cf,Creminelli:2010qf,Baumann:2011su}.
Given the promise of this type of approach, it is important to establish its regime of validity
and -- more broadly -- to what extent EFT methods are applicable in cosmology.

In this paper we take a small step toward better understanding the regime of validity of EFT as applied to inflation,
and establishing quantitative bounds for limiting the size of effects when it fails\footnote{For earlier related work see \cite{Kaloper:2003nv}.}.
We focus on the case of multi-field models in which a light field is responsible for driving inflation,
in the presence of an additional field with super-Hubble (but sub-Planckian) effective mass.
We are motivated by recent works on multi-field inflation which suggest that non-trivial field space dynamics may lead to
interesting observational signatures involving `heavy' fields
\cite{Tolley:2009fg,Cremonini:2010sv,Cremonini:2010ua,Achucarro:2010da,Baumann:2011su,Shiu:2011qw,
Chen:2009zp,Senatore:2011sp,Green:2009ds,Barnaby:2011pe,
Barnaby:2010ke, Romano:2008rr, Jackson:2011qg,Martin:2011sn,Cespedes:2012,Cook:2011hg,
Park:2012rh,Battefeld:2011yj,Peterson:2010np,Avgoustidis:2011em}
-- a result that is naively at odds with decoupling and Wilsonian thinking.
Many of these papers are reminiscent of earlier work
which explored ad-hoc features in the slow-roll potential
for possible signatures in the power-spectrum \cite{Starobinsky:1992ts,Gong:2005jr,Adams:2001vc,Chen:2006xjb}.
There it was typically found that, even with such a feature, the consistency of inflation
rendered any corrections to the power spectrum negligible.
However, for higher point functions -- and hence non-Gaussianity --
this need not be the case, making such features of potential observational interest and perhaps outside the realm of the EFT.
In this paper we attempt to clarify the notion of decoupling of a heavy field
and the regime in which the EFT description may be trusted. In particular, starting from well-motivated assumptions
on the initial conditions for the heavy field, we will bound the
size of potential EFT violations and of the resulting observational signatures.
We now summarize our main results.
\\

\noindent{\bf Summary of Main Results:}

We find that if the heavy field is initially in its vacuum at the beginning of inflation,
and its effective mass remains super-Hubble throughout inflation,
decoupling holds and the inflationary dynamics is adequately captured by EFT methods.
We make this statement quantitative by utilizing the adiabatic or WKB method to estimate the level of excitation of the heavy field,
and find that the heavy field is observationally irrelevant during a period of standard slow-roll inflation.
We discuss the case when the heavy field is initially displaced and note that such a condition seems to
require additional fine-tuning to be observationally interesting.

These results lead us to consider next an {\em ad-hoc} modification of the inflaton trajectory.
For consistency with observations, this temporary departure from inflation must be
localized in time, $\Delta t \ll H^{-1}$.
In particular, we want to ask whether a momentary violation of the slow-roll conditions can be enough to excite
a heavy field out of its initial vacuum $\langle \varphi_H \rangle =0$, potentially leading to interesting
observational signatures.
We find that such a modification can lead to a violation of decoupling (creation of quanta of the heavy field),
controlled by the dynamics of the light field and in particular by
the size of $ \vert \dot\varphi_L \dddot \varphi_L \vert $ at the onset of non-adiabaticity.
Although fine-tuned, the event must be local -- to guarantee a return to standard inflation -- and the overall process must
respect conservation of energy, even if during the event adiabaticity fails.

Using conservation of energy and requiring an appreciable amount of heavy mode production,
we establish a quantitative bound on the dynamics of the inflaton trajectory, as well as a constraint on the mass of the heavy field.
Specifically, we find that the window $\Delta t$ for particle production must satisfy a rather tight constraint,
\be
10^{-3} \left(\frac{M_{eff}}{H} \right)^{1/3} \ll \frac{ \Delta t }{H^{-1}} \ll \frac{H}{M_{eff}} \, ,
\ee
implying that adiabaticity and decoupling hold unless the effective mass of the heavy field is close to the Hubble scale,
\be
M_{eff} \lesssim 100 H \, ,
\ee
where we have been overly conservative in the last estimate.

For masses within the range
$H<M_{eff} \lesssim 100 \, H$
we make some crude estimates for the impact on the power spectrum,
with the best-case estimate giving
\be
\Delta P_\zeta \ll 10^{-5} \frac{H^2}{M_{eff} \Lambda} \, ,
\ee
suggesting that observations will be difficult in the absence of strong coupling (which here corresponds to $\Lambda \ll m_p$).
This estimate is based on the `sudden turn' approximation scheme of \cite{Shiu:2011qw}
and represents the most optimistic case, whereas other estimates lead to an even smaller effect -- a careful study of this issue
represents work in progress.
We also find that the effects of the heavy field imply a negligible change in the sound speed from unity, and therefore negligible non-Gaussianity.
Independently of our estimate, we find that the contribution to the power spectrum is controlled by the hierarchies between the Hubble scale, the heavy field effective mass and the scale of strong coupling -- a conclusion that is in agreement with the results of \cite{Baumann:2011su,Achucarro:2010da,Cremonini:2010ua} which were obtained by different methods.
We note that in the presence of strong coupling, unlike the case of DBI inflation,
here we would require a high energy completion of the two-field model.
For reasons we discuss throughout the text, this is both theoretically challenging and observationally interesting.

Our bounds can also be applied to constrain the initial conditions for models which consider
a `sharp' turn in field space.
In such models \cite{Shiu:2011qw,Achucarro:2010da,Chen:2009zp} it has been argued that allowing for a
sharp turn in the trajectory of the inflaton can lead to a potentially large signal of non-Gaussianity.
We find that typically such
scenarios require an initial displacement of the heavy field, an initial condition which is difficult to maintain for
significant amount of efoldings of inflation.
Given this, it is natural to ask what is required to generate such a displacement, under the well-motivated assumption that the heavy field
starts out in its vacuum. Our conclusion is that such a displacement
can only be generated when the effective mass of the heavy field is near the Hubble scale.

Entertaining the idea that some day such fine-tuned features in the inflaton trajectory might somehow be motivated,
we take these bounds to suggest that only for `heavy' fields near the Hubble scale one should worry about violations of EFT methods.
Of course, this is already expected, since the cutoff of the EFT should always be taken slightly
above the energy of interest for observations -- in this case $E \simeq H$.
We take this result as motivation to further understand the presence of additional massive degrees of freedom near the Hubble scale,
as well as further indication that `the EFT of Inflation' may be a robust approach to connect data with fundamental theory.
The former has been a recent line of inquiry \cite{Baumann:2011nm,Baumann:2011nk}, where it was shown that the presence of
supersymmetry may motivate additional massive fields slightly above the Hubble scale.
\\

\noindent{\bf Outline:}

The rest of the paper is organized as follows. In Section \ref{EFTmassive} we start by reviewing standard
notions of effective field theory and decoupling in cosmology.
Section III discusses the violation of adiabaticity in the model, with the corresponding production of heavy modes.
It also contains the bounds on the size of these non-adiabatic effects.
Section IV is dedicated to a discussion of work in progress, including initial estimates for the size of the effect of the production event on the
power spectrum.  In this section we also discuss the connection of our approach to the existing literature on turns in field space, as well as the 
effective field theory of inflation approach to heavy field dynamics considered in \cite{Baumann:2011su}.  Finally, the Appendix summarizes the geometric interpretation of the two field model and
gives a mapping between our notation and that used in \cite{Shiu:2011qw}.

We note that during the final stages of this work, the studies \cite{Achucarro:2012sm,Cespedes:2012hu,Jackson:2012fu} appeared which have some conceptional overlap with our discussion.  A complementary feature of our approach to those works is that our treatment of production of the heavy field can be viewed as addressing the choice of initial conditions in these papers, where the heavy field is assumed displaced from the onset.  

%%%%%%%%%
%%%%%%%%%
\section{Effective Field Theory, Massive Fields, and Cosmology
\label{EFTmassive}}
%%%%%%%%%
%%%%%%%%%

We begin with a theory described by the following two-derivative Lagrangian
\footnote{We will work with metric signature $(-,+,+,+)$ and reduced Planck mass
$m_p=1/\sqrt{8\pi G_N}=2.43 \times 10^{18}$ GeV.}
\bea
{\cal L}
&=&\half m_p^2 R- \gamma_{ab} \partial_\mu \phi^a \partial^\mu \phi^b - V(\phi) \nonumber \el
&=& \half m_p^2 R -\half \partial_\mu{\varphi}_H  \partial^\mu{\varphi}_H  -\half f(\varphi_H)
\partial_\mu{\varphi}_L  \partial^\mu{\varphi}_L - V(\varphi_H,\varphi_L) \, ,
\label{start}
\ea
where $\gamma_{ab}$ denotes the metric on field space, and for simplicity we restrict our attention to the case with only two scalar fields
$\phi^a = \{ \varphi_L,\varphi_H\}$. The notation is suggestive: $\varphi_L$ denotes a field which is light compared to the Hubble scale,
while $\varphi_H$ is a field whose effective mass is assumed to be heavy (naively $m_H>H$, however this will be made more precise below).
Given these assumptions for the hierarchy of masses, at tree level we can `integrate out' the heavy field to obtain an Effective Field Theory (EFT)
for the light field (see \cite{Burgess:2007pt} for a pedagogical review). We now outline briefly how such a procedure works.
For concreteness we take the field space metric and the potential to be
\bea
\label{fchoice}
f(\vph)&=&1+\frac{c_1}{\Lambda}\vph \, , \\
V(\vph,\vpl)&=& V(\vpl) + \half m_H^2 \, \vph^2 + h \, \vph \, \vpl^2 + \frac{c_2}{\tilde{\Lambda}} \, \vph \, \vpl^4,
\label{choices}
\eea
where $\Lambda$ and $\tilde{\Lambda}$ should be thought of as the cutoffs of the {\em two-field} theory, and
the $c_i$'s are order one constants.
Finally, we note that we are neglecting additional terms (including renormalizable interactions) only for the purpose of presentation --
although we emphasize that their inclusion is required for consistency.

With these choices and using (\ref{start}) we find the equation of motion for the heavy field is
\be
\label{GJeom}
\left( -\Box + m_H^2 \right) \vph = J(\vpl)\, ,
\ee
where $\Box\equiv -\partial_t^2 - 3H \partial_t +a^{-2} \partial_i^2$, and we have introduced the source term:
\be
\label{current}
J(\vpl)\equiv h \vpl^2   +\frac{c_1}{2 \Lambda} (\partial \vpl)^2 + \frac{c_2}{\tilde{\Lambda}}\vpl^4 \, .
\ee
Ignoring loop effects we can solve (\ref{GJeom}) formally for the heavy field and, using (\ref{choices}), we arrive at the effective action for the light field,
\be
S_{eff}=\int d^4x \, \sqrt{-g}  \left( \half m_p^2 R -\half \partial_\mu{\varphi}_L  \partial^\mu{\varphi}_L  - V(\vpl) \right)+ \Delta S \, ,
\ee
where the last term is generated from the interactions with the heavy field, and is a non-local contribution to the action:
\bea
\Delta S &=&-\half \int d^4x \, \sqrt{-g} \; J(\vpl)    \left( \frac{  1 }{\Box - m_H^2} \right)  J(\vpl).
\eea
We can then obtain a local action by expanding in powers of ${\Box}/{m_H^2}$,
\bea
\Delta S &=&\half \int d^4x \, \sqrt{-g} \; \frac{J^2(\vpl) }{m_H^2} +  \ldots \; \; . \label{correction1}
\eea
which is a good approximation when probing energies $E \ll m_H$.
The additional terms which we have dropped are suppressed by powers of ${\Box}/{m_H^2}$ and,
since we are interested in experiments or observations which probe a finite range of energies,
an appropriate number of terms can always be kept.
In the case of inflation, we will be interested in `observations' with $E \simeq H$.

Combining (\ref{current}) and (\ref{correction1}) we see that the effect of those terms is to renormalize the couplings
of the light field and generate non-renormalizable self-interactions.
As an example, the presence of the derivative interaction in (\ref{current})
generates a correction $\sim  (\partial \vpl)^4/(m_H^2 \Lambda^2)$ to the motion of the light field.
For a Friedmann-Lemaitre-Robertson-Walker (FLRW) background this leads to a correction $\sim \vpdl^4/(m_H^2 \Lambda^2)$,
which in turn gives rise to a change in the sound speed of the fluctuations \cite{ArmendarizPicon:1999rj}:
\be
c_s^2 \longrightarrow c_s^2 \simeq \frac{1}{1+\frac{8c_1^2}{m_H^2 \Lambda^2} \langle \vpdh^2 \rangle} \lesssim 1 \, .
\ee
We have introduced the notation $\langle {\cal \hat{O}} \rangle$ to denote the background value of the operator ${\cal \hat{O}}$.
Additional terms in (\ref{correction1}), loop corrections and further interactions which we have neglected for simplicity
may also lead to interesting effects in the EFT of the light field\footnote{We note that loop corrections will
be most interesting observationally when they violate a classical symmetry of the theory, i.e. they are anomalous.}.

In instances where the initial parameters of the light field are derivable from a fundamental theory (such as string theory),
it has been shown that corrections arising from integrating out the heavy field may lead to an increased likelihood for inflation
(see e.g. \cite{Dong:2010in} and references within).
However, whether this occurs is model dependent and relies sensitively on the details of the UV theory.
In fact, from the viewpoint of the EFT of the light field, as long as we are interested in energy scales below the mass of the heavy field,
observations will be relatively \emph{insensitive} to the UV physics encoded by the heavy field -- at least
{\em as long as the  EFT remains valid and we consider theories which are local and unitary} \cite{Kaloper:2002uj}.
This can be understood by looking at the parameters describing the light field (such as its mass) in the low-energy effective theory.
After the addition of appropriate counterterms to regulate divergences, these parameters may be determined by requiring a sufficient
period of inflation for the background, and by matching observations, i.e. by setting their \emph{renormalized} values equal to
those required by observation, at Hubble radius crossing\footnote{
We note that in models with $c_s<1$, modes `freeze-out' at $k\simeq aH/c_s$.
However, as discussed in \cite{Baumann:2011su}, the scaling dimension of the field must be taken into consideration -- one then finds
that the energy associated with freeze-out is still such that $E \simeq H$.
We also note that the choice of renormalization point is still arbitrary, as in the case of ordinary quantum field theory
in the absence of gravity.} $E \simeq H$.
One then finds that the effect of the corrections arising from a heavy field of mass $m_H \gg H$
will be \emph{highly suppressed}, by at least
a factor of $H^2/m_H^2$ compared to the uncorrected result \cite{Kaloper:2002uj}.
This was considered explicitly for slow-roll models of inflation in \cite{Burgess:2003zw},  where it was shown that the instances
where the heavy field could lead to observable effects were all at the cost of ruining the inflationary background.

There are at least three notable exceptions to the arguments for the smallness of the corrections discussed above.
First, it was shown in \cite{Greene:2005aj} (and references within) that for special choices of initial conditions
heavy fields could have an effect on observations of order $H/M$, with $M$ the scale of the new physics.
However, this requires assumptions about the initial state of the field which were challenged in \cite{Kaloper:2002cs},
and argued to be at odds with notions of locality and Lorentz invariance.
We will not revisit these arguments here, but instead note that such models have not yet been realized in a fundamental theory,
making a further discussion difficult.

A second way for the heavy field to have a large effect at low energy is in the case when the system becomes strongly coupled.
For example, this would be the case if the coupling in the derivative interaction in (\ref{start}) were large
(i.e. if the cutoff scale $\Lambda \ll m_p$).
A two-field model of this type, with a moderate coupling between the light and heavy field,
was considered in \cite{Tolley:2009fg}, where it was again shown that integrating out the heavy field gave rise to an effective field
theory for the light degree of freedom, with a modified sound speed $c_s \lesssim 1$.
However, as two of us discussed in \cite{Cremonini:2010ua} and was made more precise in \cite{Baumann:2011su,Shiu:2011qw},
one has to pay close attention to the hierarchy between the strong coupling scale and the mass of the heavy field.
It was found that in the class of models considered in \cite{Tolley:2009fg}, a significant deviation from $c_s \simeq 1$ required
an unnaturally strong coupling between the heavy and light field \cite{Cremonini:2010ua}.
Moreover, the strong coupling scale affected the transition to the regime of applicability of an effective description in terms
of a single field with a modified sound speed.
Of course, one must also ensure that the two-field EFT itself is valid and safe from radiative instabilities.
This makes UV-completing models with a strong coupling scale
extremely challenging, with DBI inflation being a notable example \cite{Silverstein:2003hf}.
In that case, a shift symmetry controls the derivative expansion of the model,
while a higher dimensional boost symmetry protects the model from radiative corrections.
An interesting exception to the need for strong coupling was discussed in \cite{Baumann:2011su} where it was shown that,
before reaching the strong coupling scale, there can be a reshuffling of the effective degrees of freedom of the system,
so that the theory is still described by a single field, but with a modified dispersion relation.
This was the result of the heavy field inducing a change in the scaling dimension of the light field as one approaches (but does not exceed)
the strong coupling scale.
This interesting result has further motivated studies to examine the role of additional fields with masses slightly above the Hubble scale,
which may naturally
arise from theories with spontaneously broken supersymmetry and lead to interesting observational signatures \cite{Baumann:2011nm,Baumann:2011nk}.
We should emphasize that the case of strong coupling isn't really a failure of EFT -- if quanta of the heavy field are initially present
and strongly coupled to the light field, the EFT approach was
never justified in the first place.

A third way in which a heavy field could leave an observable imprint is the case in which the EFT description simply fails.
This would be the case if the evolution of the system exhibits non-adiabaticity,  which occurs e.g.
during inflationary (p)reheating \cite{Traschen:1990sw,Kofman:1997yn}.
In the remainder of this paper we will focus on the consequences of precisely such a (temporary) violation of EFT, and
ask whether this can lead to interesting -- and observable -- effects.

%%%%%
%%%%%
\subsection{Cosmology and Regimes of Validity of the EFT}
%%%%%
%%%%%

In this section we will examine the conditions under which the decoupling of the heavy field may fail,
due to a temporary breakdown of the low energy EFT.
We will also discuss the possible effects this may have on the evolution of the early universe.
Our focus will be on cases where the EFT violation is due to a period of non-adiabatic evolution.
Thus, in order to address these issues we will start by outlining the standard formalism for defining the adiabatic vacuum
and for computing particle creation in cosmological backgrounds\footnote{For a more complete discussion we refer the reader
to the classic text \cite{Birrell:1982ix}.}.

Specializing (\ref{start}) to the case of a homogeneous and isotropic FLRW background we find:
\be
{\cal L}= \half m_p^2 R+ \half \vpdh^2 +\half f(\varphi_H) \vpdl^2 - V(\varphi_H,\varphi_L).
\ee
The resulting equations of motion are then:
\bea
\label{heom}
\ddot{\varphi}_H &+& 3H \dot{\varphi}_H - \half \partial_H f \dot{\varphi}_L^2 + \partial_H V = 0, \el
\ddot{\varphi}_L &+& 3H \dot{\varphi}_L + \partial_H (\ln f) \, \dot{\varphi}_L \dot{\varphi}_H  + f^{-1} \partial_L V = 0, \el
3H^2 m_p^2 &=& \half \vpdh^2  +\half f(\varphi_H) \vpdl^2 + V(\varphi_H,\varphi_L), \el
2\dot{H} m_p^2 &=& -\vpdh^2 - f(\varphi_H) \vpdl^2.
\label{lasthubble}
\ea
Even though by construction the heavy field won't be able to move much, $\vpddh \approx \vpdh \approx 0$,
we see from its equation of motion (\ref{heom}) that it can still have a time dependent minimum, thanks to the presence of the derivative interaction
\be
\label{Vmin}
\partial_H V =  \half \partial_H f \dot{\varphi}_L^2 \, .
\ee
Next, we want to focus on the implications of this condition.

%%%
%%%
\subsubsection{Initial displacement of the heavy field}
\label{displ}

When thinking of how to interpret (\ref{Vmin}) we should emphasize that
we are interested in the case where the light field is
related to the slow-roll parameter $\epsilon =\vpdl^2/(2 m_p^2 H^2)$ and is responsible for providing a period of inflation.
As a result, the source term $\sim \partial_H f \dot{\varphi}_L^2$ in (\ref{Vmin}) must be nearly constant, implying that
\be
\label{Vconst}
\partial_H V \sim \mbox{constant}\, .
\ee
Such a constant contribution can then be easily re-absorbed into the definition of $\vph$ (it is a tadpole),
telling us that the higher derivative interaction leads to the same physics
as a constant shift in the field.
The condition (\ref{Vconst}) translates into an initial displacement of the heavy field from its minimum,
encoded in the Lagrangian by a term such as $\sim -m_H^2\vph^{(0)}\vph$.
The fine-tuning involved in arranging for such an initial displacement -- which would be entirely {\em ad hoc} lacking a UV completion --
is then reflected in the choice of initial conditions, i.e. $\vph^{(0)}$.

Given such an initial displacement, it was shown some time ago that heavy fields can lead to oscillatory features,
with the size of
$\vph^{(0)}$ setting the amplitude of the field, and in turn
determining the observability of the effects, see e.g. \cite{Burgess:2002ub}.
In the absence of a fundamental theory to motivate the value of $\vph^{(0)}$, this seems to introduce model dependence
and fine-tuning \emph{beyond} that needed to achieve adequate inflation.
Moreover, another challenge for these models is that the oscillations of the field --
which typically scale as $\sim 1/a^3 \sim \exp(-3Ht)$ -- will be quickly damped away during inflation.
This means that for such models to have an observational impact, the displacement must survive until observable modes of the CMB are imprinted --
requiring an additional fine tuning.
Finally, in regards to the validity of the EFT, it should be emphasized that
the fact that the heavy field may have an effect does \emph{not} represent a violation of EFT methods,
since the heavy field isn't initially in its vacuum, i.e. the single field EFT was never justified.

Recently, models of this type have
received attention because of the possibility of non-Gaussian signatures
and, in many instances, for their connection to the strong coupling scale
\cite{Achucarro:2010da,Chen:2009zp, Baumann:2011su,Tolley:2009fg,Cremonini:2010ua,Cremonini:2010sv,Shiu:2011qw}.
Given this renewed interest, in this paper we want to take a critical look at the initial conditions required above,
and take an initial step towards bounding the size of possible observable effects of a heavy field.
We will address the issues of fine-tuning discussed above by
taking the heavy field to begin in its vacuum state $\langle \varphi_H \rangle = 0$.
We then ask under what conditions the background will be able to excite modes of the heavy field.
Once the heavy modes are excited, they can affect the dynamics of the light field, possibly spoiling inflation.
As we will see, the requirements that inflation is not disturbed, and that the heavy mode production happens
on a very short time scale (compared to the Hubble time) will set constraints on properties of the system.
As we have emphasized, this is very different from the case where the heavy field is assumed to be already displaced initially.

%%%%
\subsubsection{Heavy Modes and the Adiabatic Vacuum}
\label{HeavyModes}
%%%%

To determine whether modes of the heavy field that are initially in their vacuum remain there,
we begin by expanding the field in small excitations about its background value,
\beq
\delta\vph(t,\vec{x})=\vph(t,\vec{x})-\langle \vph(t) \rangle \, ,
\eq
where $\langle \vph(t) \rangle$ is the homogeneous background field which satisfies (\ref{heom})-(\ref{lasthubble}).
It is convenient to work in Fourier space with conformal time $d\eta=dt/a(t)$, and introduce a new field $\chi_k$ through
the field redefinition:
\be
\delta\vph(x,\eta)= \int \frac{d^3k}{(2 \pi)^{3/2}a(\eta)} \left( \hat{a}_k e^{i\vec{k} \cdot \vec{x}} \chi_k(\eta) +
\hat{a}^\dagger_k e^{-i\vec{k} \cdot \vec{x}} \chi^{*}_k(\eta) \right) \, .
\ee
The creation and annihilation operators obey $[\hat{a}_{k},\hat{a}^\dagger_{k^\prime}]=\delta^{(3)}(\vec{k}-\vec{k}^\prime)$,
giving rise to the field normalization condition $\chi_k \partial_\eta \chi_k^*-   \chi_k^* \partial_\eta \chi_k =i$.
Expanding the heavy field equation of motion (\ref{heom}) in fluctuations we arrive at
\be \label{feqn}
{\chi}^{\prime \prime}_k +\omega^2_k(\eta) \chi_k=0,
\ee
with primes denoting derivatives with respect to conformal time.
The frequency of the oscillator is
\be \label{ouromega}
\omega^2=k^2+a^2\left(M_{eff}^2-M_g^2\right) \, ,
\ee
and the gravitational and effective masses are given, respectively, by
\bea
M_g^2&=&\frac{1}{a^2}\left({\cal{H}}^2+{\cal{H}}^\prime\right), \nonumber \\
M_{eff}^2&=&\partial_H^2V_{eff}=- \frac{1}{2a^2} \partial^2_H f {\varphi^\prime}_L^2 + \partial^2_H V \, ,
\label{ouromegaparams}
\eea
where ${\cal{H}}=aH$ is the Hubble parameter in conformal time.
We emphasize that all terms in the frequency are evaluated on the background, but we have dropped brackets here for simplicity.
We are implicitly neglecting fluctuations of the light field and metric which can lead to rescattering effects,
but are higher order in the Hartree approximation and negligible for small occupation numbers of the heavy field.
In other words, we first seek to determine whether heavy modes can be excited at all\footnote{Such an approach is familiar from studies of
(p)reheating \cite{Kofman:1997yn}, where accounting for rescattering effects of the fields -- particularly while the `heavy field' is light -- is crucial for properly understanding the dynamics \cite{Barnaby:2009mc}.  In this paper we will focus on the question of whether heavy modes are produced leading to a violation of the light field EFT, leaving rescattering effects to future work.}.
Finally, expanding the heavy field fluctuation in positive and negative frequency modes,
\be
\chi_k=\frac{\alpha_k(\eta)}{\sqrt{2\omega}} e^{-i \int \omega d\eta} + \frac{\beta_k(\eta)}{\sqrt{2\omega}} e^{i \int \omega d\eta} \, ,
\ee
it can be shown that (\ref{feqn}) is equivalent to the two equations for the Bogoliubov coefficients
\bea
\alpha_k^\prime&=& \frac{\omega^\prime}{2 \omega} \exp \left( {2i \int \omega d\eta} \right) \beta_k, \nonumber \\
\beta_k^\prime&=& \frac{\omega^\prime}{2 \omega} \exp \left( {-2i \int \omega d\eta} \right) \alpha_k,
\eea
with the normalization condition $|\alpha|^2-|\beta|^2=1$.

To impose the condition that modes begin in their vacuum (i.e. no particles are present) one must take the boundary conditions to be
$\alpha \rightarrow 1$, $\beta \rightarrow 0$ as $k \eta \rightarrow -\infty$, indicating that there are only positive frequency modes.
Near the vacuum $\alpha$ and $\beta$ can be taken to be roughly constant and -- making use of the initial conditions --
they can be power expanded to find
\be \label{betaeqn}
\beta \approx \int d\eta \, \frac{\omega^\prime}{2 \omega} \exp \left( {-2i \int^\eta \omega(\tilde{\eta}) d\tilde{\eta}} \right),
\ee
which remains negligible as long as
\be \label{adiabconditions}
\frac{{\omega^\prime}}{\omega^2}\ll1 \, ,\bskip \frac{{\omega^{\prime \prime}}}{\omega^3}\ll1.
\ee
These equations define the adiabatic vacuum \cite{Birrell:1982ix}, and -- as long as the adiabatic conditions hold --
the heavy modes will remain frozen, allowing us to restrict our attention to the light field EFT.
If the conditions (\ref{adiabconditions}) are not respected the heavy modes will become excited,
violating decoupling. The occupation of the heavy field is then given by \cite{Kofman:1997yn}
\bea \label{chisqformula}
\langle \chi^2 \rangle &=& \frac{1}{2\pi^2 a^3(t)} \int_0^\infty dk k^2 \left| \chi_k(t) \right|^2, \nn \\
&=&\frac{1}{2\pi^2 a(t)^3} \int_0^\infty \frac{dk k^2}{\omega} \times \left( \left| \beta_k \right|^2  + Re\left( \alpha_k \beta_k^* \exp\left( -2i \int_0^\eta \omega d\eta \right) \right) \right),
\eea
where we have neglected the contribution from the Coleman-Weinberg potential, assuming (as in \cite{Kofman:1997yn})
that additional symmetries (e.g. Supersymmetry) have suppressed this contribution relative to the particle production terms.
Once the dynamics returns to an adiabatic regime (where the conditions (\ref{adiabconditions}) are again valid),
the excitations (non-zero value of $\beta$) may be interpreted as particle creation of quanta of the heavy field,
with the total number density given by
\be \label{numdensity}
n_\chi(t)=\int_0^\infty \frac{k^2 dk}{2 \pi^2 a^3(t)} \left| \beta_k \right|^2.
\ee
The energy stored in the particles produced is then given by $\rho_\chi = \omega \, n_\chi(t)$,
which for massive or non-relativistic particles becomes $\rho_\chi = m_\chi n_\chi(t)$.

%%%
%%%
\subsubsection{Breakdown of EFT and Violations of Decoupling}
%%%
%%%

The breakdown of the adiabatic approximation gives rise to excitations
of the heavy field, corresponding to a period during which decoupling can be violated.
There are a number of ways in which the adiabatic conditions may fail resulting in particle production.
A familiar example is the production of inflationary perturbations near the Hubble radius $H^{-1}$
during slow-roll inflation.
In that case the effective mass of the inflaton is much smaller than the Hubble scale, implying that it can
be neglected, and $\omega^2 \simeq k^2 -a^2 H^2$.
Thus, at the scale $k \sim aH$ the behavior of the solutions to (\ref{feqn}) change from oscillatory to a power-law instability, leading to
the production of quanta or perturbations.
Another way in which modes can be produced is if $\omega \rightarrow 0$,
implying a severe violation of the adiabatic conditions (\ref{adiabconditions}) and a production of
negative frequency modes.
Finally, we note that in order for these excitations to be interpreted as particles,
at a later time the system must re-enter the adiabatic regime (asymptotic out state),
which in the case of inflation corresponds to the exit from inflation to reheating, and a thermalized
universe\footnote{The exit from inflation to reheating defines an `out vacuum' where the notion of
particle production can be established through the adiabatic evolution as we eventually enter a radiation
dominated universe (this may take several efoldings depending on the dynamics of (p)reheating).
However, particle production is not well defined in the case of pure de Sitter space, where the notion of
an asymptotic out state is lacking. This is one reason that pure de Sitter space is not of interest for
describing the early universe inflationary period.}.

Here we are interested in the production of {\em heavy fields}.
Given our expression for the effective mass (\ref{ouromegaparams}), we can now state precisely what is meant by ``heavy,'' and
what is required to excite the field out of its vacuum state.
We expect decoupling of the heavy modes and validity of the light field EFT if and only if:
\bi
\item{initially modes begin in their vacuum state,}
\item{the adiabatic conditions (\ref{adiabconditions}) remain valid,}
\item{the effective mass in (\ref{ouromegaparams}) remains super-Hubble $M_{eff} \gg H$ for all times}
\ei
in the energy range of interest.
If all of these conditions hold, then this assures us that -- if initially in their vacuum --
the heavy modes will remain in their vacuum
and have no effect.
We again stress that if the first condition is violated -- i.e. modes
don't begin in their vacuum state but are initially displaced --
then the light field EFT was not justified in the first place.
As argued above, such initial conditions seem to require an additional
level of fine tuning beyond that required for adequate inflation.

An example of a breakdown of the adiabatic conditions has been studied
both phenomenologically in the production of superheavy dark matter \cite{Chung:2011ck},
(p)reheating \cite{Felder:1998vq,Felder:1999pv} and trapped inflation \cite{Green:2009ds},
as well as more formally in studies of the dynamics of moduli in string
compactifications \cite{Kofman:2004yc,Watson:2004aq,Cremonini:2006sx,Greene:2007sa}.
While all these studies consider initially heavy states, the effective mass of the heavy field becomes temporarily light
thanks to interactions with the light field, resulting in particle production.
Clearly, this is \emph{not} a failure of decoupling of heavy modes,
since these modes are in fact nearly massless at the time of production.
This would also be the case in the presence of a phase transition,
where one must expand around the new vacuum of the theory,
and the degrees of freedom of the new EFT can be very different.
As a simple example, we can consider a heavy scalar whose effective mass is determined by
the renormalizable interaction $g^2 \vph^2 \vpl^2$.
The heavy field dispersion relation is then given by:
\be \label{ex1}
\omega^2 = k^2 + g^2 a^2 (\vpl-\varphi_*)^2.
\ee
When $\vpl$ passes through $\vpl=\varphi_*$, the mass of the `heavy' field
vanishes, leading to a breakdown of the vacuum condition (\ref{adiabconditions}),
\be
\frac{{\omega^\prime}}{\omega^2} \sim  \frac{ {M_{eff}^\prime} }{M_{eff}^2} \sim {\cal{O}}(1),
\ee
and therefore particle production.
Although such models can be a rich source of phenomenology \cite{Barnaby:2010ke,Green:2009ds,Senatore:2011sp},
the observability of their predictions seems very sensitive to the choice of parameters of the underlying fundamental theory\footnote{From a
theoretical viewpoint these models are also interesting in that they could suggest a selection mechanism
for vacua and a new way to generate hierarchies. This is, however, beyond the scope of the present paper
\cite{Kofman:2004yc,Watson:2004aq,Cremonini:2006sx,Greene:2007sa}.}.

Instead, in the rest of this paper we will focus on the case of heavy fields whose effective mass is {\em always} super-Hubble.
There have been a number of interesting papers in the literature suggesting that even in this regime
it may be possible to have violations of decoupling and interesting observational signatures \cite{Achucarro:2010da, Chen:2009zp, Shiu:2011qw}.
As we have established, given a field that is initially in its vacuum and with super-Hubble mass (for all times),
such violations of decoupling can only come from a violation of adiabaticity.
In the next section we will consider this case and find a number of bounds on the size that such effects may have, even though the EFT approach has failed.

%%%
%%%
\section{Non-Adiabaticity During Multi-field Inflation \label{non-adiabSection}}
%%%
%%%

In this section we consider EFT violations resulting from a breakdown of the adiabatic conditions (\ref{adiabconditions}),
while requiring that the mass of the heavy field is always strictly super-Hubble, $M_{eff} > H$.
Our goal is to obtain a quantitative estimate for the maximal size of such violations and what this implies both
for the decoupling of the heavy field and the evolution of the cosmological background.

In the previous section we reviewed the connection between the presence of time dependent masses (and therefore frequencies) in cosmological backgrounds and exciting heavy modes out of their vacuum state.
If this time dependence is strong enough heavy modes will be produced during the period of non-adiabaticity leading to a failure of
the light field EFT description.
Recall from (\ref{feqn}) that the fluctuations of the heavy field obey
\be \label{feqnagain}
{\varphi}^{\prime \prime}_H +\left(   k^2+a^2m_H^2- \frac{1}{2} \langle \partial^2_H f {\varphi^\prime}_L^2\rangle -a^2M_g^2 \right) \varphi_H=0,
\ee
where $M_g^2=({\cal{H}}^2+{\cal{H}}^\prime)/a^2$ and we have defined
$m_H^2 \equiv \langle \partial^2_H V \rangle=M_{eff}^2+\frac{1}{2} \langle \partial^2_H f {\varphi^\prime}_L^2\rangle$.
We again use $\langle {\cal \hat{O}} \rangle$ to denote the background value of an operator ${\cal \hat{O}}$, and in this section we will adopt the notation that both the real and Fourier transformed components of the heavy field fluctuations are denoted by $\vph$.

There are three possible sources of non-adiabaticity appearing in (\ref{feqnagain}).
One is the gravitational mass $M_g(t)$ (i.e. the evolution of the cosmological background),
the other two are the contributions of the dynamics of the light and heavy field to the effective mass term $M_{eff}(t)$.
If any of these give a significant contribution to the adiabatic parameters
$\frac{{\omega^\prime}}{\omega^2}$ and $\frac{{\omega^{\prime \prime}}}{\omega^3}$,
excitations will be produced and decoupling violated.

\subsection{Gravitational Sources of Violation \label{GRproduction}}
We start by considering the time dependence arising from the gravitational mass term $M_g$ in the heavy field fluctuation equation (\ref{feqnagain}).
By assumption (since the heavy field is taken to be initially in its vacuum, $\langle \vph \rangle =0$)
we are interested in an inflationary background driven by the light scalar, so we require the dynamics of
$\varphi_L$ to be such that $\epsilon \equiv \frac{d}{dt}(H^{-1}) \ll 1$.
With this assumption,
we can immediately see that there will not be any significant contribution to non-adiabaticity arising
purely from the gravitational sector.
In fact, we have
\be \label{adiabgravity}
\frac{{\omega^\prime}}{\omega^2} \simeq \frac{(a^2 M_{eff}^2)^\prime}{2 \omega^3}
\simeq \frac{H}{M_{eff}} \, ,
\ee
where we have kept only the leading terms and neglected the time dependence in $M_{eff}$ to focus on gravitational effects.
By construction we have ${H}/{M_{eff}}<1$, telling us that the gravitational background during a period of inflation is \emph{not} a significant
source of particle production of massive quanta -- decoupling of the massive field prevails.

We note that one might be concerned that we have ignored the gravitational terms $M_g^2 \sim H^2$ in (\ref{feqnagain}).
However, the associated non-adiabaticity is easily seen to be negligible since
$\dot{\omega}/\omega^2 \sim \dot{M}_g/M_g^2 \sim \dot{H}/H^2 = -\epsilon$, and for inflation $\epsilon \ll 1$.
It is for the same reason that one can not rely on purely gravitational reheating of the universe after inflation.
Of course, if we are not interested in an inflationary background this conclusion could change, depending on the particular cosmological history.
However, for power-law expansion with $a\sim t^p$ the non-adiabaticity is at most $\sim 1/p$, and one finds negligible production of
massive fields -- as expected for an `adiabatically' expanding universe.
In conclusion, we find that significant violations of decoupling \emph{from gravitational effects alone}
would require a strong violation of the slow-roll conditions and a significant departure from inflation.

\subsection{Non-gravitational Sources of Violation -- A First Look \label{particleprod}}
We have just seen that during inflation gravitational effects alone are not enough to invalidate standard arguments for
decoupling of the massive field.
We now turn to the question of whether non-gravitational effects -- associated with the dynamics of the light field --
can lead to significant particle production.
As a warm up exercise and to establish our approach, in this section we begin by neglecting the derivative interaction in (\ref{feqnagain}),
and consider instead the time dependence arising from an interaction term of the form $g^2 \vpl^2 \vph^2$.
We will return to the original expression for the frequency (\ref{feqnagain}) shortly.
Since our interest is in dynamics acting on scales which are small compared to gravity, i.e. $\Delta t \ll H^{-1}$,
we can neglect the cosmic expansion and take the limit in which gravity decouples, i.e. $m_p \rightarrow \infty$ and $ H \rightarrow 0$ ($a=1$) so that $\epsilon \sim \vpdl^2 / (H^2 m_p^2)$ remains fixed.
The relevant dispersion relation for fluctuations of the heavy field is then:
\be \label{gswex1}
\omega^2=k^2 +m_H^2 + g^2 \vpl^2.
\ee
We should note that, unlike in the example of moduli trapping (\ref{ex1}), here we always have $M_{eff}^2=m_H^2+g^2 \vpl^2 > H^2$.
We now want to determine whether  (\ref{gswex1}) can lead to a large enough source of non-adiabaticity to violate decoupling.

To address this question we will utilize a formalism developed in \cite{Chung:1998bt} (see also \cite{Lawrence:1995ct,Gubser:2003vk})
for estimating the amount of particle production -- i.e. the size of (\ref{betaeqn}) -- using the method of steepest descent.
We are interested in dispersion relations of the form
\be \label{dansomega}
\omega^2=k^2+m_0^2 \, C(\eta),
\ee
where $m_0^2$ is a constant parameter, $C(\eta)$ encodes all of the time dependence (gravitational or otherwise),
and therefore $M(\eta)\equiv m_0 \, C^{1/2}$ acts as a time-dependent mass for the heavy field.
At the onset of non-adiabaticity $\omega$ becomes complex, signifying an instability towards particle production,
with the peak of the production occurring at a time which we will denote by $\tau = r + i\mu$.
The use of the steepest descent method is valid as long as the moment of production is sharply peaked
relative to the other time scales in the problem -- a condition which must checked in each case.
Once we have determined $\mu$ and $r$, the amount of particle production per mode can then be estimated as
\beq
\label{PPk}
|\beta_k|^2 \simeq
e^{-\pi \mu \, \omega(r)}.
\eq

We are now ready to apply this estimate to (\ref{gswex1}).
Let's assume first that the light field has negligible acceleration, so without loss of generality we can let $\vpl=vt$,
with $v$ the initial velocity (for now we will ignore the fact that the light field is indeed the inflaton).
Using this parametrization in (\ref{gswex1}) and setting $\omega^2=0$, we find that the peak of production corresponds to
a time\footnote{Since we are working in the limit in which gravity decouples ($a=1$ and $H=0$),
there is no need to distinguish between conformal and coordinate time.} $t^2=-(k^2+m_H^2)/(gv)^2$,
telling us that $\mu=\sqrt{k^2+m_H^2}/(gv)$ and $r=0$.
Combining these expressions with (\ref{gswex1}) and (\ref{PPk}) we arrive at
\be \label{gswex1beta}
|\beta_k|^2 \simeq
\exp\left[-\pi \left( \frac{k^2+m_H^2}{gv}  \right)\right] \, ,
\ee
which reproduces the results of \cite{Kofman:2004yc,Watson:2004aq} in the limit in which the heavy field mass comes only
from the interaction (i.e., $m_H=0$).  However, here we can see that the mass $m_H$ leads to an additional suppression of $|\beta_k|^2$,
as expected.  We emphasize that the computational `trick' of identifying the poles $\omega^2=0$ in the complex plane
is to extend the WKB wave function to include the effects of particle production,
and that the frequency never vanishes in real space\footnote{A related, but alternative,
approach based on quantum mechanical scattering appears in \cite{Kofman:1997yn} and yields the same results we obtain here.}.

Next, we will apply this result to the case in which the light field is responsible for driving inflation.
We now have $\vpdl=\sqrt{2\epsilon}Hm_P$, and so we identify $v=\sqrt{2\epsilon}Hm_P$.
Moreover, the interaction coupling $g$ must be weak enough not to interfere with the inflationary expansion.
From (\ref{gswex1beta}) we see that the exponential suppression is now controlled by
$\sim m_H^2 / (g\sqrt{\epsilon}Hm_P)\sim  (m_H/H)^2 / (10^{5}g)$,
where we have made use of the COBE normalization $(H/m_p)^2 \simeq 10^{-10} \epsilon$.
For required choices of the coupling $g \ll 1$ and for fields with mass $m_H \gg H$ we find once again that the heavy field decouples.
However, this result could have been easily anticipated, since the light field drives inflation and therefore dominates the energy density
of the system -- its dynamics is by definition gravitational, and the argument of Sect. \ref{GRproduction} applies.

Finally, one might worry that the inclusion of the derivative interaction in (\ref{feqnagain}) would change this conclusion.
However, it does not.
If the heavy field is initially in its vacuum,  $\partial_H^2 f$ must be constant.
Taking the scale of this term to be $\sim 1/\Lambda^2$, we see that the derivative interaction in (\ref{feqnagain}) scales as
\be
\frac{1}{2} \langle \partial^2_H f {\dot{\varphi}}_L^2\rangle \sim \frac{{\dot{\varphi}}_L^2}{\Lambda^2}
\sim \frac{\epsilon H^2 m_p^2}{\Lambda^2}\sim \epsilon H^2 \ll m_H^2,
\ee
which is negligible, assuming $\Lambda$ is not too far below the Planck scale.  Thus, we arrive at one of our main results:
{\em the only way to violate decoupling of the heavy field is with a departure from inflation}.
We will make this point more explicit in the next section.
An important caveat is if the model lies in a regime of severe strong coupling so that $\Lambda \ll m_p$.
As we have previously discussed, examples of this kind were studied in
\cite{Tolley:2009fg,Cremonini:2010ua,Cremonini:2010sv,Shiu:2011qw} and for cases near but not exceeding strong coupling in
\cite{Baumann:2011su}. For the former possibility -- in the absence of a UV completion -- the origin of such a low scale,
the issue of additional non-renormalizable operators suppressed by this scale, and the radiative stability of
the {\em two field} model remain an important challenge.

%%%%
%%%%
\subsection{Non-gravitational Sources of Violation -- Departure from Inflation \label{violsection}}
%%%%
%%%%

We have seen that significant production of the heavy field requires a departure from inflation and a modification
of the dynamics of the light field.
Given this, we want to repeat the calculations of the previous section allowing now for an arbitrary behavior of the light field,
to establish what is required to obtain a significant violation of decoupling.
In particular, we want to ask whether a momentary violation of the inflationary conditions can be enough to
excite the heavy field out of its initial vacuum $\langle \phi_h \rangle =0$, and perhaps lead to interesting observational signatures.
A temporary departure from inflation can be achieved by introducing either a sharp feature in the potential or significant curvature
in the scalar field space.  This will then alter the trajectory of the inflaton, which may result in a significant amount of non-adiabaticity.

As before, we are interested in dispersion relations of the form
\beq
\label{frequency}
\omega^2 = k^2 +m_0^2C(t),
\eq
where $m_0^2$ is a constant parameter and for now $C(t)$ is completely arbitrary and encodes any time dependence which may appear
(e.g. from interactions or gravitational effects).  However, an important restriction we need to impose on $C(t)$ is the requirement
that the inflationary conditions are satisfied
\emph{on long time scales} -- this will force any violations to be local relative to a Hubble time, $\Delta t \ll H^{-1}$.
This will again imply that gravitational effects will be negligible (allowing us to set $a=1, H=0$).

\begin{figure*}[t]
\centering \label{fig1}
\includegraphics[scale=0.40]{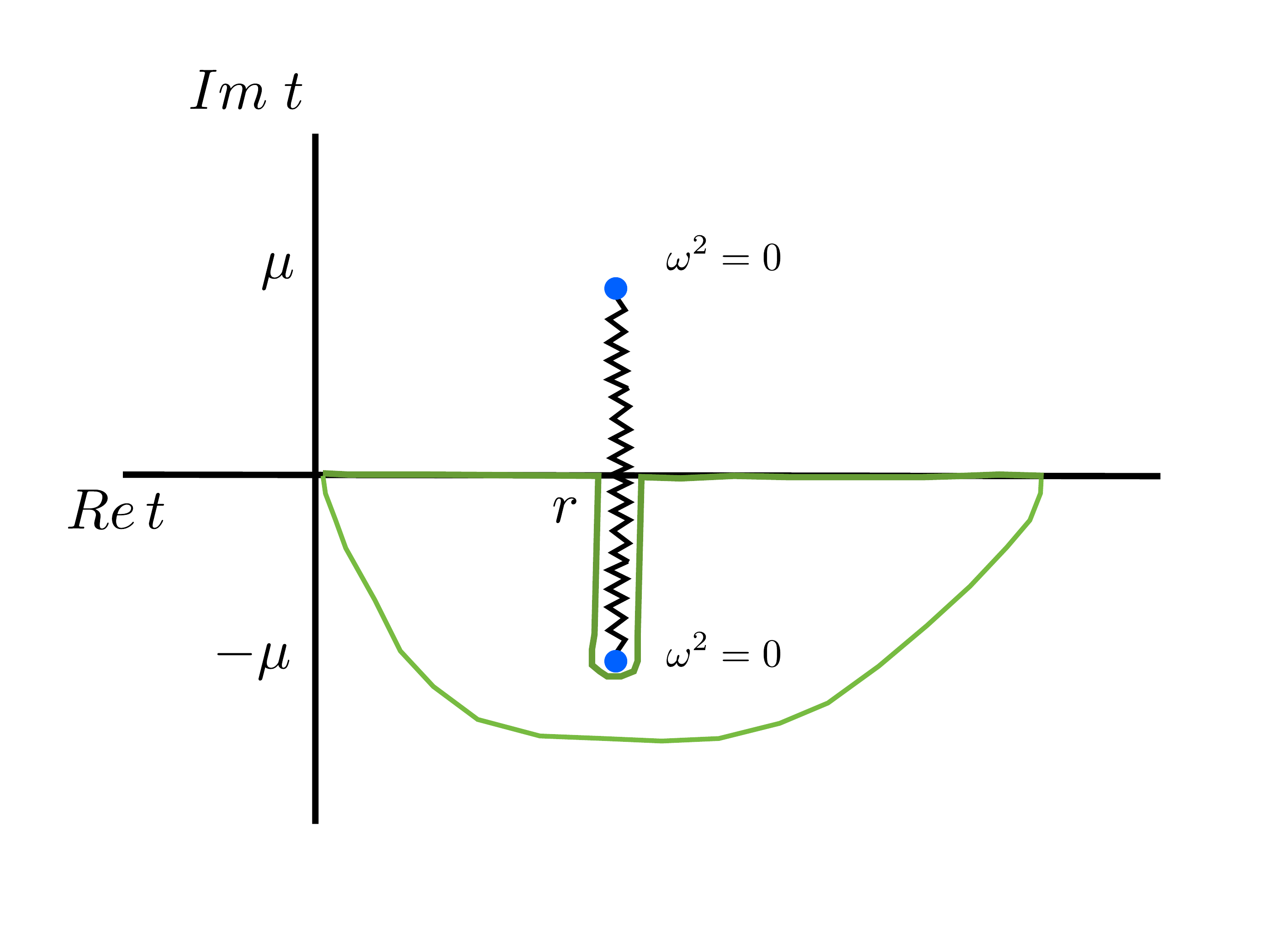}
\caption{Non-adiabaticity becomes appreciable for times $t \gtrsim r$ and is peaked at the pole where $\omega^2=0$ and $t=\tau=r+i\mu$.
At early and late times adiabaticity holds and time is strictly real. }
\end{figure*}

Recall that the estimate (\ref{PPk}) for violations of adiabaticity requires finding the time $\tau=r+i\mu$ at which the production is peaked.
To obtain an explicit expression for $r$ and $\mu$ following \cite{Chung:1998bt} it turns out to be convenient to
expand $C(t)$ around the onset of non-adiabaticity,
where the time (and frequency) is analytically continued to the complex plane.
The peak of the non-adiabaticity then corresponds to poles where the frequency vanishes, which in this case is given by $C(\tau)=-k^2/m_0^2$.
Expanding $C(t)$ about $t_*=r+i0$ then gives:
\bea \label{expandc}
C(t)=  C(r)  + \dot{C}(r)   (t-r)  + \half \ddot{C}(r) (t-r)^2 + {\cal O} (t-r)^3.
\eea
Evaluated at the time of maximum production $\tau=r+i\mu$, the expression takes the form:
\be \label{expandthing}
C(\tau)= C(r) +i \dot{C}(r)\mu-\half \ddot{C}(r)\mu^2+ {\cal O} (t-t_\ast)^3. \\
\ee
Finally, equating this expansion with $C(\tau)=-k^2/m_0^2$ yields the two conditions
\bea \label{condition1s}
C(r)-\half \ddot{C}(r)\mu^2&=&-\frac{k^2}{m_0^2}, \\
\dot{C}(r) \mu&=&0, \label{condition2s}
\eea
for the real and imaginary parts, respectively, and we note that we have dropped subleading terms.
From the latter condition we find $\dot{C}(r)=0$ (since $\mu=0$ would correspond to no production) implying that $r$ is
the location on the real axis of an extremum\footnote{If the period of non-adiabaticity were prolonged, higher order terms in the expansion
(\ref{expandthing}) may become important and this could shift the extrema. However, here we are interested in rapid violations with $t-t_*=i\mu$ small.}.
Moreover, using (\ref{condition1s}) along with the dispersion relation evaluated at $r$ we find
\be
\half \ddot{C}(r)\mu^2=\frac{k^2}{m_0^2}+C(r)=\frac{\omega(r)^2}{m_0^2},
\ee
implying that $\ddot{C}(r)>0$ and so $r$ is indeed the location on the real axis at which the effective mass $m_0^2 C(t)$ is minimized.
Finally, combining all the ingredients above we arrive at an estimate for the amount of particle production associated with an arbitrary
function $C(t)$:
\bea \label{estbeta}
|\beta_k|^2 &\simeq& e^{-\pi \mu \, \omega(r)}, \\
&\simeq& \exp \left( -\pi \left[    \frac{k^2+m_0^2C(r)}{m_0 \sqrt{\ddot{C}(r)}}
\right]  \right)=  \exp \left(-\frac{\pi \omega^2(r)}{m_0 \sqrt{\ddot{C}(r)}}    \right) \, .
\eea

As a simple check of our computation, we can revisit the case of moduli trapping of a massive field, which we discussed in Sect. \ref{particleprod}.
From (\ref{gswex1}) we identify $m_0\equiv m_H$ and $C(t)=1+(g\varphi_L)^2/m_H^2$, and recall that we had chosen $\varphi_L = v t$.
It is then straightforward to check that  $C(r)=1$ and
$\ddot{C}(r)=2 g^2 v^2 / m_H^2$, giving
\bea \label{result1}
|\beta_k|^2 &\simeq&
\exp\left[-\pi \left( \frac{k^2+m_H^2}{gv}  \right)\right] \, ,
\eea
in agreement with our earlier result (\ref{gswex1beta}).

Now that we have outlined the method for computing the amount of non-adiabaticity, and hence the possibility of violations of decoupling
of the heavy field, we are ready to tackle the derivative interaction in (\ref{feqnagain}).
In this case the term $m_0^2 \, C(t)$ which plays the role of an effective mass is controlled by
\beq
\label{ourC}
C(t) \equiv 1-\frac{1}{2m_H^2}\, \langle \p_H^2 f \, \dot{\vp}_L^2 (t) \rangle,
\eq
and once again we have made the identification $m_0\equiv m_H$.
It is important to note that, since the heavy field \emph{evaluated on the background} is assumed to vanish initially,
i.e. $\langle \phi_H \rangle = 0$, the term $\langle \p_H^2 f \rangle$ does not change in time (initially) and
is therefore constant.
This would not be the case if we had displaced the heavy field initially.
However,  as emphasized earlier in the paper, preserving such a condition
during inflation would require additional fine-tuning and violate the validity of the light field EFT from the start.
Thus, the only source of time dependence in the effective mass term (\ref{ourC}) must come from the evolution of the
light field $\langle \dot{\vp}^2_L (t) \rangle$.
In particular, the kinetic energy of the light field will be transferred into the effective mass of the heavy field,
lowering its size, and therefore leading to the possibility of production.
Moreover, as we already discussed, requiring that
this gives rise to a significant amount of particle production
entails building in a feature, to allow for a temporary departure from inflation.

Following the formalism outlined above for estimating production, we expand $C(t)$ around the time $t_\ast=r+i0$ at which non-adiabaticity begins,
and we find
\bea \label{cr1}
\left. C \right(r)&=& 1-  \frac{1}{2m_H^2}\, \left. \langle \p_H^2 f \, \dot{\vp}_L^2  \rangle \right\vert_{r} \; ,\\
\ddot{C}(r) &=&
- \frac{1}{m_H^2} \left. \langle \p_H^2 f \left( \ddot{\vp}^2_L  + \dot{\vp}_L \dddot{\vp}_L \right) \rangle \right \vert_{r} \,.
\ea
The conditions (\ref{condition1s}) and (\ref{condition2s}) imply that $\ddot{\vp}_L(r)=0$ and
\beq \label{crdd1}
\ddot{C}(r)  =
- \frac{1}{m_H^2} \left. \langle \p_H^2 f \left(  \dot{\vp}_L \dddot{\vp}_L \right) \rangle \right \vert_{r} >0 \, .
\eq
As we will discuss shortly, we are interested in the case $\langle \p_H^2 f \rangle >0 $, and therefore we must require that
$\left. \langle \dot{\vp}_L \dddot{\vp}_L \rangle \right \vert_{r}<0$.
In other words, there must be a sudden jolt in the behavior of the light field (an abrupt change in its acceleration) in order
to create quanta of the heavy field.
Using these results in the estimate (\ref{estbeta}) we find
\be
\label{beta}
|\beta_k|^2 \simeq
\exp\left(
-\pi \, \frac{ k^2+m_H^2 -\half  \langle \p_H^2 f \, \dot{\vp}_L^2  \rangle}
 {\sqrt{ \left| \langle \p_H^2f   (  \dot{\vp}_L \dddot{\vp}_L) \rangle \right| }  }
  \right),
\ee
where all quantities inside the exponential are to be evaluated at $t=r$.
Let's comment briefly on the structure of this result.
First of all, notice that the higher the mass $m_H$ of the heavy field, the smaller the amount of particle production, as expected.
Next, the effect of the term $ \langle \p_H^2 f \, \dot{\vp}_L^2  \rangle$ in the numerator is to counteract the suppression due to $m_H$:
the field space curvature and the evolution of the light field (in particular, its velocity $\dot{\vp}_L$) allow for an
\emph{enhancement} of particle production.
This is nothing but a reflection of the transfer of energy from the light field kinetic energy to the heavy field sector.
Finally, from the denominator we see that the larger the change in the acceleration of the light field $\sim \p_t\ddot{\vp}_L$, the larger the particle production.

From (\ref{beta}) we also see that the time scale associated with the production is\footnote{We note that another way of
estimating the time scale is to utilize the uncertainty principle as was discussed in \cite{Kofman:2004yc}.
It was shown there that the uncertainty principle implied a time scale of
production $\Delta t \sim 1/\sqrt{gv}$.  That result can be obtained here by noting that the argument of the
exponential in (\ref{result1}) is parametrically of the form $\omega^2(r) \Delta t^2$, allowing us to find an estimate
in agreement with \cite{Kofman:2004yc}.}
\beq \label{timeinterval}
(\Delta t)^4  =\frac{1}{\langle  \partial_H^2 f \dot{\varphi}_L(r)\dddot{\varphi}_L(r) \rangle }
\sim  \frac{ \Lambda^2}{\langle   \dot{\varphi}_L(r)\dddot{\varphi}_L(r) \rangle},
\eq
which gives the intuitive result that the duration of time production (or non-adiabaticity)
decreases the stronger the coupling (the smaller $\Lambda$ is) and the stronger the change in the acceleration (the large $\vpdddl$ is).
We emphasize that the time scale for appreciable production $\Delta t$ is much less than the previously found parameter $\mu=t-t_*$.
This is because at the beginning and end of the period of non-adiabaticity
there is little
contribution from the production of particles, so that significant production only occurs during an interval $\Delta t  \ll 2\mu$.

\subsection{Bounds on the Size of the Violation}

We start with the observation that if we want to ensure that the source of non-adiabaticity
leads to a significant violation of decoupling, the exponential
in (\ref{estbeta}) should not be heavily suppressed.
This leads to the constraint
\be
   \frac{k^2+m_H^2 \, C(r)}{m_H \sqrt{\ddot{C}(r)}}  \ll1.
\ee
Note that production will favor long wavelength modes (although sub-Hubble), so we can neglect the wave number $k$ in the expression above
compared to the mass term. Using (\ref{cr1}) and (\ref{crdd1}) we arrive at the constraint
\bea
\label{con1}
m_H C(r) &\ll& \sqrt{\ddot{C}(r)} \simeq
\frac{ \sqrt{ \left| \vpdl(r) \vpdddl(r) \right|}}{m_H\Lambda}\, ,
\eea
where we have taken\footnote{Recall that $\langle \vph \rangle=0$, which implies that $\langle \partial_H^2f \rangle$ is constant
up to the moment violation begins, at $t=r$.} the constant value of $\langle \partial_H^2f \rangle \sim \Lambda^{-2}$.
This constraint can be simplified by noting that $C(r)$ is by definition the ratio $M_{eff}^2/m_H^2$ at the onset of production.
Thus, we arrive at the following lower bound for the quantity $\left| \vpdl(r) \vpdddl(r) \right| / \Lambda^2$
which controls the amount of non-adiabaticity
\be
\label{mconstraint1}
\frac{  \left| \vpdl(r) \vpdddl(r) \right|}{\Lambda^2 M_{eff}^{4}} \gg 1 \,.
\ee

An upper bound on this quantity can be obtained by making use of energy conservation.
Recall that we are assuming that the particle production event is local, in the sense that $\Delta t \ll H^{-1}$.
Thus, we require that the Hubble parameter is left unchanged by the period of non-adiabaticity.
Since during inflation $3 H^2 m_p^2 \simeq V$, this means that the source of energy for the transition must come entirely
from the kinetic term of the inflaton.
If all of its energy went into producing heavy particles with energy density $\rho_H$,
conservation of energy would imply $\rho_H < \vpdl^2/2$, where before the event we have $\vpdl^2 = 2\epsilon H^2 m_p^2$.
In practice we must require some kinetic energy of the light field to remain so that inflation can end and so we have
the upper bound $\rho_H \ll \vpdl^2$.

Since the excitations of the heavy field will be massive the corresponding energy density will scale like pressure-less
matter with $\rho_H = m(t) n_H(t)$,
where $n_H(t)$ is given by (\ref{numdensity}).
We can again neglect the diluting effect of the expansion and set $a=1$ in (\ref{numdensity}),
since we are interested in conserving energy at the moment of production.
We find that the number density is
\be
\label{numberdensity}
n_H \simeq \left( m_0 \sqrt{\ddot{C}(r)} \right)^{3/2} \exp \left(-\pi \frac{m_0 C(r)}{\sqrt{\ddot{C}(r)}} \right)
\simeq  \left( m_0 \sqrt{\ddot{C}(r)} \right)^{3/2} \, , \nonumber \\
\ee
where in the last step we have considered the case of interest where production is not highly suppressed by the exponential.
\begin{figure*}[t]
\centering \label{fig2}
\includegraphics[scale=1.00]{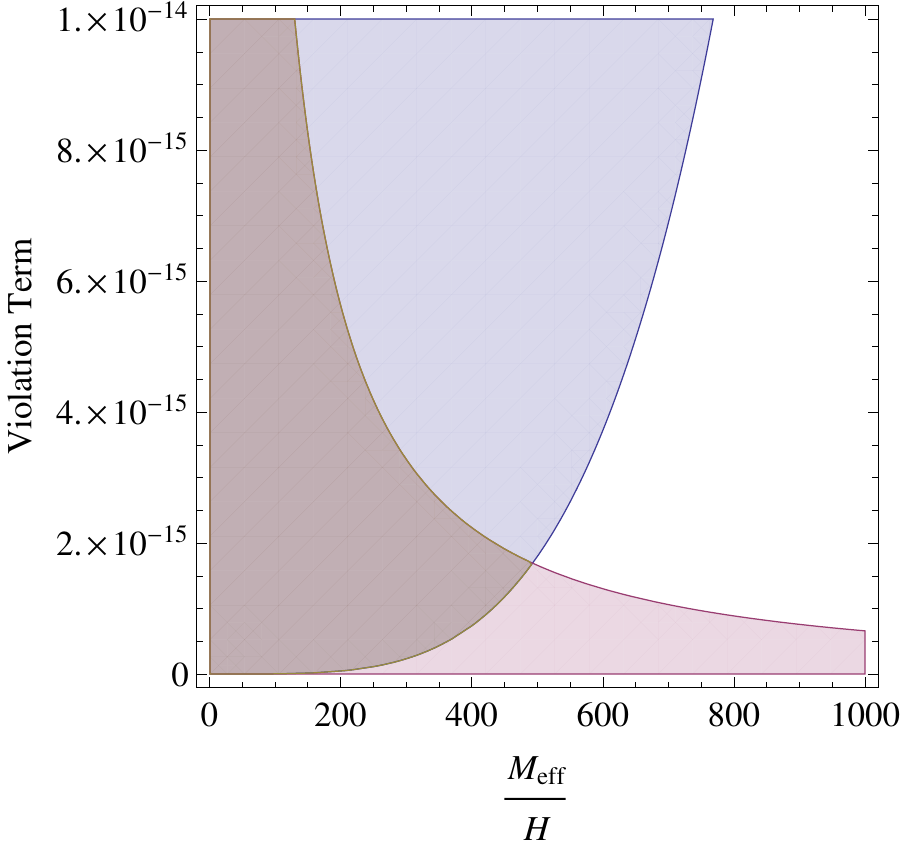}
\caption{Constraints on the level of non-adiabaticity and production of super-Hubble mass particles during a
temporary violation of inflation.  The dark shaded area above represents the allowed region of production where
the constraints overlap.  The upper bound (blue) comes from demanding conservation of energy during the violation,
whereas the lower bound (red) comes from requiring a large enough violation so that modes are produced.
Although our estimates indicate a sharp cutoff in production around $M_{eff}=500 H$ as seen in the graph,
this result is in fact overly optimistic given that the regions actually saturate the lower and upper bound.
More realistically we must be deep within these regions so that $M_{eff} \simeq H$ and so even with a violation
of the EFT only for masses near the Hubble scale are models interesting.}
\end{figure*}
The energy density slightly after production is then
\be
\rho_H \simeq m_0 \, C^{1/2}(r)  \left( m_0 \sqrt{\ddot{C}(r)} \right)^{3/2},
\ee
and will then become diluted once inflation resumes as $\rho_H \rightarrow \rho_H / a^3 \sim \rho_H e^{-3Ht} $.
Using (\ref{crdd1}) and the definition of $C(r)$ we find that requiring $\rho_H$ to be much less than the kinetic energy of the inflaton
leads to the following upper bound
\bea
M_{eff}\left( m_H \sqrt{\frac{\vpdl(r) \vpdddl(r)}{\Lambda^2 m_H^2}} \right)^{3/2} & \ll & \epsilon H^2 m_p^2 \, . \nonumber
\eea
Using the COBE normalization $\epsilon \simeq 10^{10} H^2/m_p^2$ we have\footnote{We note that we are interested
in the situation where the net change in the slow-roll parameter will be negligible, although during the transition
slow-roll will be violated. This is the case because we require inflation to resume after the violation,
with little change in the Hubble scale and scalar potential as stressed above.}
\be
\frac{  \left| \vpdl(r) \vpdddl(r) \right|}{\Lambda^2 M_{eff}^4} \ll 10^{13} \left( \frac{H}{M_{eff}} \right)^{16/3} \, .
\ee

Combining this result with the previous lower bound (\ref{mconstraint1}) we find
\be \label{mainresult}
1 \ll  \frac{  \left| \vpdl(r) \vpdddl(r) \right|}{\Lambda^2 M_{eff}^4} \ll 10^{13} \left( \frac{H}{M_{eff}} \right)^{16/3} \, ,
\ee
implying a rather tight window for whatever dynamics of the light field is responsible for driving the period of non-adiabaticity.
We see that the upper bound, although quite easily satisfied for masses near the Hubble scale,
quickly drops off as we increase the effective mass of the heavy field (assumed to be above the Hubble scale);
the window for particle production gets tighter and tighter, and the
bound begins to tightly constrain the possibility of a violation.
Considering the unrealistic case when the upper bound is saturated and all energy from the kinetic term of the light field went into production, with the
fiducial values of $H=10^{12}$ GeV, $\Lambda = m_p$, and $\epsilon=10^{-2}$ we find that for masses much greater than $100H$ the heavy modes can not become excited
-- even allowing for departures from adiabaticity and inflation.
This can also be seen in Figure 2.
We emphasize that the bound was derived by making very conservative estimates, and a more accurate estimate would push the mass
even closer to the Hubble scale.

We can use (\ref{mainresult}) to get a bound on the time interval during which production takes place, a quantity that is perhaps more intuitive.
Using (\ref{timeinterval}) we find the window
\be
10^{-3} \left( \frac{M_{eff}}{ H} \right)^{4/3} M_{eff}^{-1} \ll \Delta t \ll M_{eff}^{-1}\, ,
\ee
where the upper bound demonstrates that for non-adiabaticity and production the time scale must be shorter than the
Compton Wavelength ($\lambda_c \sim M_{eff}^{-1}$), and the lower bound again follows from conservation of energy.
We note that these bounds are more stringent than the simple requirement that the production acts on scales which are decoupled from gravity, i.e. $\Delta t \ll H^{-1}$.
This can be seen more clearly by re-expressing the bound in terms of Hubble time as
\be
10^{-3} \left(\frac{M_{eff}}{H} \right)^{1/3} \ll \frac{ \Delta t }{H^{-1}} \ll \frac{H}{M_{eff}},
\ee
demonstrating the narrow range for which production is possible.
For example, with an effective mass of $M_{eff} \simeq 100 H$, particle production becomes very hard to achieve -- heavy modes remain
entirely decoupled even though we allowed for a potential EFT violation.

%%%%%
%%%%%
\section{Observational Implications, Connection to Recent Studies, and Future Directions \label{additionalinteractions}}
%%%%%
%%%%%
In Section \ref{non-adiabSection} we studied the production of the heavy field and found that significant production required an
ad-hoc modification of the inflaton trajectory.
Here we estimate the size of the effect on the power spectrum resulting
from such a modification while imposing the bounds we derived in the last section. We also discuss how our setup relates to multi-field inflation
models which contain sharp-turns in field space, as well as to the recent work of \cite{Baumann:2011su}
which utilizes the {\em Effective Field Theory of Inflation}~\cite{Cheung:2007st} to understand the observable effects of changing
the scaling dimension of the light field.

\subsection{Power Spectrum Estimate}
To estimate the power spectrum we begin by introducing perturbations for the light and heavy fields.
Denoting the metric perturbation by $\psi=\psi(t,\vec{x})$, the gauge-invariant Mukhanov-Sasaki variables are given by
\bea
Q_H&=& \delta \varphi_H + \frac{\vpdh}{H}\psi \, , \\
Q_L&=& \delta \varphi_L + \frac{{\vpdl}}{H}\psi \, .
\eea
In spatially flat gauge ($\psi=0$) these correspond to fluctuations of the heavy and light field, respectively.
The equations of motion for the $\{Q_H,Q_L\}$ perturbations are quite involved and can be found e.g. in \cite{Lalak:2007vi}.
However, it is possible to redefine the perturbations so that one field corresponds
to fluctuations along the trajectory of the inflaton (adiabatic mode), while the other to fluctuations perpendicular to the trajectory
(isocurvature mode) \cite{Gordon:2000hv}.
This can be achieved by performing a rotation in field space, with the rotation angle sensitive to the amount of kinetic energy stored
in the heavy field, i.e. to the term ${\dot\varphi_H} /  {\sqrt{\dot\varphi_H^2 + f \dot\varphi_L^2}}$.
In particular, when the heavy field kinetic term is subdominant, $\dot\varphi_H^2 \ll f \dot\varphi_L^2$,
the corresponding rotation angle $\dt$ is very small,
\beq
\label{rotationangle}
\dt \simeq \frac{\dot\varphi_H}{\sqrt{f \dot\varphi_L^2}} \ll 1\, .
\eq
In this case the relationship between the original perturbations and the adiabatic and isocurvature fluctuations is particularly simple,
\bea
\left(\begin{array}{c}
 v_H \\ v_L
\end{array} \right)=
\left(\begin{array}{cc}
-1 & \;\;\;\; {\dt} \sqrt{f}  \\
\dt & \sqrt{f}
\end{array} \right)
\left(\begin{array}{c}
 a Q_H \\ a Q_L
\end{array} \right) \, ,
\label{newperts}
\eea
with $\dt$ parametrizing the amount of mixing between the `light' and `heavy'
modes\footnote{In the notation of \cite{DiMarco:2005nq,Lalak:2007vi}, the rotation angle $\theta$ was defined via
$\cos\theta = \frac{\dot\varphi_H}{\sqrt{\dot\varphi_H^2 + f \dot\varphi_L^2}}$.
Here we are focusing on the case where $\dot\varphi_H^2 \ll f \dot\varphi_L^2$, for which $\theta \sim \pi/2$.
Thus, for this particular case our $\dt$ is related to $\theta$ through $\theta = \delta\theta-\pi/2$.}.
This transformation also acts to canonically normalize the modes so that their quantization is straightforward.

Generically, the couplings between the curvature and isocurvature fluctuations are controlled both by the heavy field dynamics
and by the behavior of the light field.
Working in the limit in which gravity decouples
and taking for simplicity a separable potential,
the quadratic action for the perturbations reduces to the simple form
\be \label{act1}
S=-\int d^4x \, \frac{1}{2} \Bigg[    \dot{v}_H^2 - \left( k^2 + \cM_H^2 \right) v_H^2  + \dot{v}_L^2 - \left( k^2 + \cM_L^2 \right) v_L^2   \Bigg]
+ S_{mix},
\ee
where the mixing terms for the perturbations are given by
\be
S_{mix}=\int d^4x \, \Bigg[ - \xi \,
%\frac{\partial_Hf}{\sqrt{f}} \vpdl
v_L \dot{v}_H + \cM_{LH} \, v_L v_H  \Bigg] \, ,
\ee
%\be
%S_{mix}=\int d^4x \, \Bigg[ \frac{\partial_Hf}{\sqrt{f}} \vpdl
%\dot{v}_L v_H + \cM_{HL} v_L v_H  \Bigg] \, ,
%\ee
and the various couplings by
\begin{eqnarray}
\cM_{H}^2 &=&  \partial_H^2V- \frac{1}{2} \partial^2_H f \vpdl^2 + \ldots \; ,\\
\cM_{L}^2 &=& f^{-1} \partial_L^2 V  + \ldots \; , \\
\xi &=&  \frac{\partial_H f}{\sqrt{f}} \dot\varphi_L  + \ldots \;, \\
\cM_{LH} &=&  - \dot\xi \, ,
%- f^{-1/2} \partial_Hf \vpddl  + \ldots \; ,\; .
\end{eqnarray}
where the neglected terms are ${\cal O} (\delta\theta, \dot{\delta\theta})$ which we have omitted for simplicity.
In fact, given the requirement that the mass of the heavy field {\em always} satisfies $M_{eff} > H$,
here we would like to focus on the effect of the light field only, and will therefore assume that $\dot\varphi_H \simeq \ddot\varphi_H \simeq 0$,
effectively switching off the heavy field dynamics encoded in the $\dt$ terms\footnote{Although it is clear that $\dt \ll 1$,
this does not guarantee that the contributions coming from the $\dt$ terms (e.g.
terms of the form $\frac{\ddot\varphi_L}{\dot\varphi_L} \dt$ in the kinetic mixing term)
will in fact be subdominant. A complete analysis would have to incorporate such terms, and we leave it to future work.}.
The equations of motion are then
\bea
\ddot{v}_H &+& \left( k^2 + \cM^2_H \right) v_H  +  \frac{\partial_Hf}{\sqrt{f}} \vpdl \dot{v}_L  =0 \, , \\
\ddot{v}_L &+& \left( k^2 + \cM^2_L \right) v_L - \frac{\partial_Hf}{\sqrt{f}} \vpdl \dot{v}_H
% + \cM_{LH} \,
- \frac{\partial_Hf}{\sqrt{f}} \vpddl v_H=0 \, ,
\eea
%%%
%\bea
%\ddot{v}_H &+& \left( k^2 + \cM^2_H \right) v_H  +  \frac{\partial_Hf}{\sqrt{f}} \vpdl \dot{v}_L + \cM_{HL} \, v_L =0 \, , \\
%\ddot{v}_L &+& \left( k^2 + \cM^2_L \right) v_L - \frac{\partial_Hf}{\sqrt{f}} \vpdl \dot{v}_H+ \cM_{LH} \, v_H=0 \, ,
%\eea
where we note that metric perturbations are absent due to taking the decoupling limit.
Indeed, we have emphasized throughout the text that the light field feature which gives rise to particle production
must be localized on a very small time scale
$\Delta t \ll H^{-1}$.
At the end of the period of non-adiabaticity, the light field will return
to its standard slow-roll behavior, and the heavy field will be quickly damped away by the inflationary expansion.
In this paper we will estimate the size of the effect on the power spectrum by treating this event as instantaneous, as was done recently in
\cite{Shiu:2011qw} for the case of a sharp turn in field space\footnote{For another recent
attempt to develop an analytic understanding of sharp features in the inflaton trajectory see \cite{Adshead:2011jq}.}.
Thus, on long time scales  -- i.e. for the purpose of estimating the power spectrum -- the feature in the light field dynamics
can be taken to be well approximated by a delta function
localized at some time $t_\ast$, i.e.
\beq
\label{deltafnapprox}
\dot\varphi_L (t) \simeq \Delta \varphi_L  \, \delta(t-t_\ast) = - H \, \Delta \varphi_L \, \eta_\ast \, \delta(\eta-\eta_\ast) \, ,
\eq
where an expression for $\Delta\varphi_L$ can be obtained by expanding the light field around the time of production (\ref{scalarserieslast}),
\beq
\Delta\varphi_L = \int \dot\varphi_L(t) \, dt = \dot\varphi_L(r) \, \Delta t + {\cal O}(\Delta t^2 )\, .
\eq

Given this approximation to capture the temporally localized event we can now restore the (long-wavelength) dependence on the gravitational background.
The equations of motion for the perturbations then take the form
\begin{eqnarray}
\label{heavypert}
\partial^2_\eta v_L &+& \left(k^2-\frac{2}{\eta^2}\right) v_L + \frac{1}{H \eta} \partial_\eta \left(\frac{\partial_Hf}{\sqrt{f} } \dot\varphi_L   \, v_H\right)
- \frac{2}{H \eta^2} \frac{\partial_Hf}{\sqrt{f}}\dot\varphi_L v_H = 0 \, ,\\
\label{lightpert}
\partial^2_\eta v_H &+& \left(k^2-\frac{2}{\eta^2}+\frac{{M}_{eff}^2}{H^2 \eta^2}\right) v_H
- \frac{\partial_Hf\dot\varphi_L}{  \sqrt{f} \, \eta H } \, \partial_\eta v_L - \frac{\partial_Hf\dot\varphi_L}{\sqrt{f} \, \eta^2 H }
v_L = 0 \, ,
\end{eqnarray}
where we have neglected the mass of the light field as it is small relative to the leading terms.
Using these equations and following the approach of \cite{Shiu:2011qw} we obtain an estimate for the power spectrum
resulting from the event
\bea
P_{\zeta} = P_\zeta^0+\Delta P_\zeta &\simeq& \frac{H^2}{8\pi^2 m_p^2\epsilon} \left[1 + \frac{\partial_H f}{\sqrt{f}} \, \Delta \varphi_L  \, \sin(2k\eta_\ast) \right] \nn \\
&\simeq & \frac{H^2}{8\pi^2 m_p^2 \epsilon} \left[1 + \frac{\dot\varphi_L(r)}{\Lambda} \,  \Delta t  \, \sin\left(\frac{2k}{k_\ast}\right) \right]  \, ,
\ea
where in the last line we have made use of $k_\ast \eta_\ast \equiv -1$ and $\frac{\partial_H f}{\sqrt{f}} \sim \frac{1}{\Lambda}$,
and defined the power spectrum in the absence of a violation to be $P_\zeta^{0}$.
In particular, we note that
\beq
\frac{\Delta P_\zeta }{P_\zeta^{0}}= \frac{\partial_H f}{\sqrt{f}} \, \dot\varphi_L(r) \, \Delta t
\sim \frac{\dot\varphi_L(r)}{\Lambda^{1/2} (\dot\varphi_L  \dddot\varphi_L)^{1/4}}  \ll \sqrt{\epsilon} \frac{H}{M_{eff}} \frac{m_p}{\Lambda} \, ,
\eq
where we have imposed the lower bound on the level of adiabaticity (\ref{mainresult}).
Using the COBE normalization $P_\zeta^{0} \simeq 10^{-10}$ we find that our best-case estimate for the correction obeys
\be \label{bigbound}
\Delta P_\zeta \ll 10^{-5} \frac{H^2}{M_{eff} \Lambda}.
\ee
This suggests that such an effect would be difficult to observe
in the absence of strong coupling, which here corresponds to $\Lambda \ll m_p$.

We emphasize again that this result -- based on the approximation scheme of \cite{Shiu:2011qw} -- represents our most
promising estimate for observations and other estimates suggest an even small effect. For example, an underlying assumption
in the approach of \cite{Shiu:2011qw} is that
$v_H$ and $v_L$ are maximally correlated, which requires sub-Hubble physics we refrained
from specifying.  Had the perturbations $v_L$ and $v_H$ been the usually assumed quantum
fluctuations of two different fields, they would be statistically independent and their contributions should be added to the power spectrum in quadrature.
This leads to a correction significantly smaller than the bound above, strengthening the conclusion that the
effect is difficult to probe with observations.

However, this view may be too pessimistic.  In particular, a more realistic calculation of the power spectrum would require
understanding the behavior of the fields during production, a process which although local in time is intrinsically not classical --
e.g. the notion of particle number of the heavy quanta is not well defined during the event (only after).  This is reminiscent of the situation during inflationary
preheating where lattice techniques are often valuable.  Another strategy to obtaining a more accurate estimate of the power spectrum
would be to further develop the analysis of \cite{Green:2009ds}, where similar obstacles were encountered.
Finally, although the contribution to the power spectrum is small, a better understanding of interactions during the
production event may lead to interesting features in the bi-spectrum or higher point functions.  However, our preliminary estimates suggest
a bound for the sound speed of the light field model, and the corresponding level of non-gaussianity $f_{NL} \sim 1/c_s^2$.
Using the bound (\ref{bigbound}) and comparing to the power spectrum we see that the corresponding speed of sound will be bounded by
\bea
\frac{1-c_s}{c_s} \simeq \Delta P_\zeta / P_\zeta^0 \ll 10^{5} \frac{H^2}{M_{eff} \Lambda}.
\eea
For a hierarchy of scales corresponding to $H< M_{eff}$ and $H< \Lambda$, with the latter required to avoid
the perturbative breakdown of the two-field model, we find negligible variation of the sound speed from $c_s \simeq 1$
and non-gaussianity $f_{NL} \sim 1/c_s^2 \simeq 1$.  This result for the sound speed and its dependence on the
strong coupling scale agrees with \cite{Cremonini:2010ua,Baumann:2011su,Shiu:2011qw,Leblond:2008gg},
whose authors used very different methods.

Independently of our estimate, we find that the contribution to the power spectrum is controlled by the hierarchies between the Hubble scale,
the heavy field effective mass and the scale of strong coupling -- a conclusion that is in agreement with the results of
\cite{Baumann:2011su,Achucarro:2010da,Cremonini:2010ua}.
We note that in the presence of strong coupling, unlike the case of DBI inflation,
here we would require a high energy completion of the two-field model.
A careful study of these issues represents work in progress.

%%%%%%%%%%%%%%%%%%%%%%%%%%%%%%%%%%%
\subsection{Connection to the Geometric Approach to Multi-field Inflation}
In this section we would like to briefly discuss how our model relates to recent studies of multi-field models
of inflation, where dynamics may result from the non-trivial geometry of the scalar
field space \cite{Peterson:2010np,Cremonini:2010sv,Cremonini:2010ua,Shiu:2011qw,
Achucarro:2010da,Chen:2009zp}.  Additional details on the relationship between our approach and geometric methods for
the field space can be found in the Appendix.

We begin by recalling that the original light and heavy field fluctuations are related to curvature and isocurvature perturbations
by the rotation (\ref{newperts}), controlled by $\dt \simeq \frac{\dot\varphi_H}{\sqrt{f \dot\varphi_L^2}}$.
This description makes the physical interpretation manifest -- the misalignment between the two sets of perturbations
is a measure of how dynamical the heavy field is.
If $\varphi_H$ were exactly at rest, the two sets of fluctuations would remain aligned as the system evolves.
A turn in the inflaton trajectory (as discussed in the references above)
is one example of how such a misalignment $\dt \neq 0$ may occur.
In this paper we are interested in cases where the heavy field always has a super-Hubble scale mass.
Thus, although in (\ref{newperts}) we allow for a slight turn during the period of violation of adiabaticity $\dt \neq 0$,
we expect the hierarchy of masses to keep this angle small.
In fact, we can estimate the size of $\dt$ by using the fact that $\half \dot\varphi_H^2 \lesssim \rho_H$,
with $\rho_H$ the energy density of the heavy modes produced during the period of non-adiabaticity.
Given that the energy for production of the heavy modes comes from the light field's kinetic energy
(so that the Hubble scale is roughly the same before and after inflation) we have
 $\rho_H \ll \sqrt{f} \dot\varphi_L^2$.  Thus, we have
\beq
\dt
\lesssim \frac{\sqrt{\rho_H}}{\sqrt{f}\dot\varphi_L} \ll 1 \, ,
\eq
confirming the intuition that the rotation angle should be small.
On the other hand, for models in which the `light' and `heavy' perturbations can exchange roles,
as discussed in \cite{Cremonini:2010ua, Shiu:2011qw},
the rotation angle could be quite large and the approximation (\ref{newperts}) would not hold.
We emphasize that here we are not interested in this case, since it does not constitute a violation of decoupling
but simply the emergence of light degrees of freedom -- i.e., the single field EFT is no longer appropriate (see also \cite{Baumann:2011su}).

Following our assumption in this paper that the heavy field begins in its vacuum,
the inflaton trajectory is initially straight.
However, once the heavy field is excited,
one has the possibility of \emph{generating} a turn in the trajectory.
More precisely, denoting by $T^a$ and $N^a$ the vectors tangent and normal to the curve,
the turning parameter $\eta_\perp$ is defined
in terms of the rate of change of the tangent vector,
\beq
\frac{DT^a}{dt} = - H \eta_\perp N^a \, .
\label{Dt}
\eq
A non-zero turning parameter is associated with a displacement of the heavy field from its minimum, along
the direction normal to the trajectory.
To see this explicitly we note that (\ref{Dt}), combined with the equation of motion along the trajectory in field space (\ref{sigmaeom}),
leads to the expression
\beq
\label{etaperpVN}
\eta_\perp = \frac{V_N}{H \sqrt{\dot\varphi_H^2 + f \dot\varphi_L^2} } \, ,
\eq
where $V_N = N^a V_a$ is the projection of the gradient of the potential along the normal direction.
Thus, by allowing for heavy field production and a non-zero $\eta_\perp$, we have generated the initial heavy field displacement which we discussed in Section \ref{displ}.

In particular, for the case of negligible dynamics of the heavy field $\dot\varphi_H \simeq \ddot\varphi_H \simeq 0$
(the $\dt=0$ case on which our power spectrum estimate was based),
the heavy field background equation of motion (\ref{heom}) reduces to
\beq
\label{backgroundEMO}
\partial_H V \simeq \half \partial_H f \dot{\varphi}_L^2 \, .
\eq
This is precisely of the form of (\ref{etaperpVN}) and describes a shift from the minimum
controlled by a
(constant)
turn rate given by
\beq
\eta_\perp = \half \frac{\p_H f \dot\varphi_L}{\sqrt{f} H} ,
\eq
where the light field is evaluated at the peak of particle production.
Finally, we note that the correction to the power spectrum can also be expressed in terms of $\eta_\perp$,
\beq
\frac{\Delta P_\zeta}{P_\zeta^0} \simeq  2 H \eta_\perp \Delta t \ll 2 \eta_\perp \frac{H}{M_{eff}} \, ,
\eq
where we made use of the bound on the window for particle production we derived in the last section.
This provides a bound on the size of the effect that sharp turns may have on the power spectrum, which were recently
studied in \cite{Peterson:2010np,Cremonini:2010sv,Cremonini:2010ua,Shiu:2011qw,
Achucarro:2010da,Chen:2009zp}.  A further mapping of our
approach to that of \cite{Shiu:2011qw} is presented in the Appendix.

%%%%%%%%%%%%%%%%%%%%%%%%%%%%%%%%%%%%%%%%%%%%%%%%%%%%
%%%%%%%%%%%%%%%%%%%%%%%%%%%%%%%%%%%%%%%%%%%%%%%%%%%%
\subsection{Connection to work of Baumann and Green and the EFT of Inflation}
We would like to close this section by making contact with the analysis of \cite{Baumann:2011su}, who restricted
their attention to the situation in which the light field obeys a shift symmetry.
In this case, terms in the action proportional to $v_L^2$ and $v_L v_H$ are forbidden, and the equations of motion for the fluctuations reduce to
\bea
\label{shiftEOMs}
\ddot{v}_H &+& \left( k^2 + \cM^2_H \right) v_H  +
\frac{\partial_Hf}{\sqrt{f}} \vpdl
\dot{v}_L  =0 \, , \\
\ddot{v}_L &+& k^2 \, v_L - \frac{\partial_Hf}{\sqrt{f}} \vpdl
\dot{v}_H = 0 \, ,
\eea
where we recall that $\cM^2_H =  M^2_{eff}$.

To better understand the behavior of these two coupled oscillators, it is convenient to look at their normal modes.
Going to momentum space, the equations of motion become
\bea
\left(\begin{array}{cc}
\omega^2-{k^2} - \cM_H^2  &
i\omega \xi  \\
-i\omega \xi &
\omega^2-{k^2}
\end{array} \right)
\left(\begin{array}{c} v_H \\v_L \end{array}\right)
=0,
\eea
where $\xi \equiv \frac{\partial_Hf}{\sqrt{f}} \vpdl$.
The dispersion relations are then found from the determinant to be
\be \label{disp}
\omega^2_{\pm} = k^2 + \half \left( \cM_H^2  + \xi^2 \right) \pm \
\sqrt{\xi^2 k^2 + \frac{\left( \cM_H^2 + \xi^2 \right)^2}{4}} \, .
\ee
We see that for momenta in the range $k^2 \gg \xi^2 \gg \cM_H^2$,  (\ref{disp}) describes
two massless modes with equal frequencies $\omega \sim k$.
On the other hand, when
 we are interested in momenta $k^2 \ll \xi^2$ the frequencies can be approximated
 by\footnote{Here, in the $\dt=0$ case this condition is equivalent to $k^2 \ll \epsilon \frac{H^2 M_P^2}{\Lambda^2}$.}
\beq
\label{omegaexp}
\omega^2_{\pm} = k^2 + \half \left( \cM_H^2  + \xi^2 \right) \pm \left[\half \left( \cM_H^2  + \xi^2 \right)
+ \frac{\xi^2 k^2}{\cM_H^2  + \xi^2} - \frac{\xi^4 k^4}{(\cM_H^2  + \xi^2)^3} + \ldots \right] \, . \nn
\eq
We can now identify a massive mode with $\omega_+^2 \sim \cM_H^2  + \xi^2$, and a light degree of freedom with
\beq
\omega_-^2 \simeq k^2\left(1-\frac{\xi^2}{\cM_H^2 + \xi^2}\right) + {\cal O}\left(\frac{k^4}{\xi^2}\right) = c_s^2 k^2,
\eq
from which we can immediately read off the speed of sound:
\beq
c_s^2 = \frac{\cM_H^2}{\cM_H^2 + \xi^2} \, .
\eq
Here we see clearly that the role of the coupling $\xi$
%\propto \frac{\partial_H f}{\sqrt{f}}\dot\varphi_L$
is to push the sound speed below one.
However, to bring $c_s$ very close to zero requires $\xi^2 \gg \cM_H^2$ which, in the  $\dt=0$ approximation,
translates into the condition
\beq
M_{eff} \ll \sqrt{\epsilon} H \frac{M_P}{\Lambda} \simeq 10^5 \frac{H^2}{\Lambda} \, ,
\eq
where in the last line we have used the COBE normalization and again used that ${\cal M}_H \equiv M_{eff}$.
Clearly, this condition is very hard to satisfy
for super-Hubble masses $\cM_{eff} > H$ unless the coupling scale $\Lambda$ is taken to be
extremely low, as already emphasized in \cite{Cremonini:2010ua}.
Note also that one recovers the standard $c_s \simeq 1$ result by increasing the size of the effective mass, as expected.

Finally, we note that when $c_s \, \xi < k <\xi$ the term that dominates in the expansion (\ref{omegaexp}) of the light degree of freedom
is the quartic term,
\beq
\omega_-^2 \simeq \frac{\xi^4 k^4}{(\cM_H^2  + \xi^2)^3} \simeq \frac{k^4}{\xi^2} \, ,
\eq
leading to the change in dispersion relation observed in \cite{Baumann:2011su} for modes with energy
$c_s^2 \, \xi < \omega < \xi$.
The frequency at which the dispersion relation is modified sets the scale of new physics, which in our case (for $\dt=0$) can be estimated to be
\beq
\omega_{new} \equiv c_s^2 \xi \simeq c_s^2 \frac{\sqrt{\epsilon} H M_P}{\Lambda} \simeq 10^5 \, c_s^2 \, \frac{H^2}{\Lambda} \, .
\eq
and so the scale of new physics relative to the heavy field effective mass is
\beq
\frac{\omega_{new}}{M_{eff}} \simeq 10^5 \, c_s^2 \, \left(\frac{H^2}{\Lambda M_{eff}}\right) \, ,
\eq
again demonstrating the balance between the scales of the heavy field ($M_{eff}$), gravity ($H$), and the coupling $\Lambda$.
We see that for high scale inflation $H\simeq 10^{12}$ GeV and with $\Lambda \simeq m_p$ implying $c_s \simeq 1$,
and the scale of new physics is just below the mass of the heavy field as expected. However, for other choices of the hierarchy there are
narrow windows where $\omega_{new}$ can be below the scale of strong coupling and the modified dispersion relation
gives interesting physics.  This agrees with the results
of \cite{Baumann:2011su} where the EFT of inflation and Goldstone approach were instead used.

\section*{Acknowledgments}
We would like to thank Neil Barnaby, Daniel Baumann, Cora Dvorkin, Dan Green, Nemanja Kaloper, Sarah Shandera and Gary Shiu for useful discussions.
This work is supported in part by STFC.
The work of A.A. was supported by a CTC Postdoctoral Fellowship at DAMTP, University of
Cambridge and in part by a Marie Curie IEF Fellowship at the the University of Nottingham.
S.C. is grateful to KITP for hospitality during the workshop on Holographic Duality and Condensed Matter Physics
while this work was underway. The work of S.C. has been supported by the Cambridge-Mitchell Collaboration in Theoretical
Cosmology, and the Mitchell Family Foundation.
R.H.R. is supported by Funda\c{c}\~{a}o para a Ci\^{e}ncia e a Tecnologia (Portugal) through the grant SFRH/BD/35984/2007,
and acknowledges a research scholarship by the Cambridge Philosophical Society.
K.T. is partly supported by the MNiSW grants N N202 091839 and IP2011 056971.
S.W. is supported by the Syracuse University College of Arts and Sciences.
%%%%%%%%%%%%%%%%%%%%%%%% Begin Appendix %%%%%%%%%
%%%%%%%%%%%%%%%%%%%%%%%%%%%%%%%%%%%%%%%%%%%%%%%%%
\appendix
%%%%%%%%%%%%%%%%%%%%%%%%%%%%%%%%%%%%%%%%

\section{Geometric Interpretation of the Two Field Model \label{app:curvature}}
We begin this appendix with a brief review of how the physics of the two field model is understood within the geometry of the field space
and the trajectory of the inflaton. This approach to analyzing multi-field inflation models was first advocated in \cite{GrootNibbelink:2001qt},
and was discussed more recently in \cite{Peterson:2010np,Shiu:2011qw,Achucarro:2010da,Peterson:2011yt}.
After a brief review, we give an explicit map between our parameters -- keeping explicit the dependence on the light and heavy fields --
and those that appeared recently in \cite{Shiu:2011qw,Achucarro:2010da} in the context of the geometric approach.

\subsection{Basics of Curves in \Rthree}
We can parametrize an arbitrary curve in \Rthree as
\be
\vec{\alpha}(t)=\left(x(t),y(t),z(t) \right) \, .
\ee
The tangent to the curve is then
\be
T^a=  \frac{\dot{\vec{\alpha}}}{\left| \dot{\vec{\alpha}}  \right|}=\frac{\left( \dot{x}, \dot{y}, \dot{z} \right)}{\sqrt{\dot{x}^2 + \dot{y}^2 +\dot{z}^2}} ,
\ee
where we have defined $\dot{x}\equiv \frac{dx}{dt}$.
Although this gives the tangent vector in terms of coordinate time, we are usually interested in the arc
length, $ds = \left| \dot{\vec{\alpha}}  \right| dt$, which allows us to write $T=T(s)$.
Then the intrinsic or geodesic curvature is defined as
\be
\kappa = \left| \frac{dT}{ds} \right| = \frac{1}{| \dot{\vec{\alpha}}|} \frac{dT}{dt} \, .
\ee
Note that for a curve with constant curvature, this takes the simple form
\be
\kappa= \left| \frac{dT}{ds} \right| = \frac{1}{R} \, ,
\ee
with $R$ the radius of curvature.
The normal to the curve is instead given by
\be
N^a=\frac{1}{\kappa} \, \frac{dT^a}{ds} = \frac{1}{\kappa | \dot{\vec{\alpha}}|} \, \frac{dT^a}{dt} \, ,
\ee
whereas the binormal is $B=T \times N$, with
\be
\frac{dB^a}{ds}=- \tau N^a \, ,
\ee
where $\tau$ is the torsion.  The way in which the  tangent, normal, and binormal (or {\it Frenet Frame})
change along the curve are summarized by
\bea
\frac{d}{dt} \left(\begin{array}{c}T \\N \\B\end{array}\right)=
| \dot{\vec{\alpha}}| \left(\begin{array}{ccc} 0& \kappa  &0  \\-\kappa  & 0 & \tau \\0 & - \tau  & 0\end{array}\right)
 \left(\begin{array}{c}T \\N \\B\end{array}\right),
\eea
with their change in \Rthree given by the curvature and torsion.
Note that they obey the following properties:
\bea
T^aT_a=N^aN_a=B^aB_a=1 \, , \el
T^aN_a=T^aB_a=N^aB_a=0 \, .
\eea

\subsection{Geometry of the Field Space}
We are interested in using the basic geometric notions of the previous section to parametrize the physics of the field space.
The Lagrangian we are going to be considering is
\bea
\frac{{\cal L}}{\sqrt{-g}} &=&- \gamma_{ab} \partial_\mu \phi^a \partial^\mu \phi^b - V(\phi), \el
&=&  \half (\dot{\varphi}_H)^2  +\half f(\varphi_H) (\dot{\varphi}_L)^2 - V(\varphi_H,\varphi_L) \, ,
\eea
where we have restricted our attention to the case of two fields, and have allowed for
a non-trivial field space metric, (i.e. $\gamma_{ab} \neq \delta_{ab}$).
The resulting equations of motion are
\bea \label{eoms}
\ddot{\varphi}_H &+& 3H \dot{\varphi}_H - \half \partial_H f \dot{\varphi}_L^2 + \partial_H V = 0, \el
\ddot{\varphi}_L &+& 3H \dot{\varphi}_L + \partial_H (\ln f) \, \dot{\varphi}_L \dot{\varphi}_H  + f^{-1} \partial_L V = 0, \el
3H^2 m_p^2 &=& \half (\dot{\varphi}_H)^2  +\half f(\varphi_H) (\dot{\varphi}_L)^2 + V(\varphi_H,\varphi_L), \el
2\dot{H} m_p^2 &=& -(\dot{\varphi}_H)^2 - f(\varphi_H) (\dot{\varphi}_L)^2.
\eea
The definitions of the last section carry over, but we must introduce a new time
derivative $D_t$
which acts as a kind of covariant derivative:
\be
D_t X^a=\frac{d}{dt} X^a + \Gamma^a_{bc} X^b \dot{\phi}^c.
\ee
Our formulae then remain valid, with this derivative in place of the
normal time derivative.  For example, in terms of a generic metric on field space $\gamma_{ab}$ the curvature becomes
\be
\kappa = \frac{1}{| \dot{\vec{\alpha}}|} \left| D_tT \right| =\frac{1}{| \dot{\vec{\alpha}}|}  \sqrt{\gamma_{ab} D_tT^a D_tT^b } \, .
\ee
Geometric quantities for our field space metric (curvature is in units of inverse mass) are then given by the following
expressions,
\bea
\Gamma^H_{LL}&=&- \half \partial_H f \bskip \Gamma^L_{LH}=\Gamma^L_{HL}=\half \partial_H (\ln f), \el
R&=&\frac{1}{2f^2} \left[ (\partial_H f)^2 -2 f \partial_H^2 f \right]\, ,
\label{extcurvature}
\eea
where we note that the latter is the extrinsic curvature and should not be confused with the intrinsic curvature $\kappa$.

Introducing the definition $\dot{\sigma} \equiv \left| \dot{\vec{\alpha}} \right|$ we have
\be \label{sigmadef}
\dot{\sigma}^2 = \dot{\varphi}_H^2 + f \dot{\varphi}_L^2 \, .
\ee
The tangent vector components are
\bea
T^1=\frac{\dot{\varphi}_H}{\dot{\sigma}}, \bskip T^2=\frac{\dot{\varphi}_L}{\dot{\sigma}},
\eea
where the index one and two refer to the heavy and light field, respectively.
The normal vector components instead are given by:
\bea
N^1=-\frac{1}{V_N}\left( \vpddh-\half \partial_Hf \, \vpdl^2- \frac{\sdd}{\sd} \dot{\varphi}_H \right), \el
N^2=-\frac{1}{V_N}\left( \vpddl+\partial_H(\ln f) \, \vpdh \vpdl - \frac{\sdd}{\sd} \vpdl \right) \, ,
\eea
where $V_N$ is the projection of the gradient potential along the normal direction, $V_N=V_a N^a$.

The equation of motion along the trajectory in field space is
\be \label{sigmaeom}
\ddot{\sigma}+3H\sd+ V_\sigma=0 \, ,
\ee
where $V_\sigma = V_a T^a$ is the potential projected along the tangent direction.
The Einstein equations may also be written in terms of the inflaton motion along the curve:
\bea
3H^2 m_p^2 &=& \half \sd^2 + V, \el
2\dot{H} m_p^2 &=& -\sd^2.
\eea
The slow roll conditions then become
\bea
\epsilon&=&-\frac{\dot{H}}{H^2}=\frac{\sd^2}{2 m_p^2 H^2}, \\
\eta^a&=&-\frac{1}{\sd H} D_t \dot{\varphi}^a,
\eea
and using
\bea \label{using}
D_t \vpdh&=&\vpddh-\half \partial_Hf \, \vpdl^2, \el
D_t \vpdl&=&\vpddl+ \partial_H(\ln f) \, \vpdl \vpdh,
\label{using2}
\eea
we have:
\bea
\eta^1&=&-\frac{1}{\sd H} \left( \vpddh-\half \partial_Hf \, \vpdl^2 \right), \\
\eta^2&=&-\frac{1}{\sd H} \left( \vpddl+ \partial_H(\ln f) \, \vpdl \vpdh \right).
\eea

From these expressions one can show that the curvature is
\bea \label{intcurvature}
\kappa &=&\frac{1}{\sd^2} \sqrt{\left(  D_t\vpdh - \frac{\sdd}{\sd}
    \vpdh \right)^2 + f(\varphi_H) \left( D_t\vpdl - \frac{\sdd}{\sd}
    \vpdl \right)^2} \nn \el
&=&\sqrt{\frac{f(\vph)}{(\vpdh^2+f(\vph)\vpdl^2)^3}}\left| \vpddl
  \vpdh-\vpddh \vpdl + \partial_H(\ln f) \, \vpdh^2 \vpdl+\half
  \partial_Hf \vpdl^3  \right|,
\eea
It is important to emphasize that although this expression
involves non-zero curvature of the field space (via the function
$f(\varphi_H)$ -- see (\ref{extcurvature}) above) this doesn't necessarily
guarantee curvature of the trajectory (\ref{intcurvature}).
For example, we can have non-zero field space curvature while having $\kappa=0$.
We can also obtain a non-zero $\kappa$, even if the field space curvature
completely vanishes (e.g., $\gamma_{ab}=\delta_{ab}$).  This is because
particular choices of the parameters in the potential can lead to
curved trajectories as well.

Since we are interested in local effects of the scalar field (meaning $\Delta t \ll H^{-1}$),
gravity decouples and the essential physics is simply described in terms of the curvature of the trajectory.  However, it is possible to
define a slow turning parameter $\eta_\perp$ as was done e.g. in \cite{Achucarro:2010da, Shiu:2011qw} and using our definition of
curvature we find
\be
\left| \eta_\perp \right|=\frac{\sd}{H} \kappa= \frac{f^{1/2}}{\sd^2 H} \left| \vpddl
  \vpdh-\vpddh \vpdl + \partial_H(\ln f) \, \vpdh^2 \vpdl+\half
  \partial_Hf \vpdl^3  \right| \, .
\ee
Note that in the limit in which $\dot\varphi_H$ and $\ddot\varphi_H$ can be neglected,
the turning parameter reduces to
\be
\left| \eta_\perp \right|=\frac{\sd}{H} \kappa \simeq \frac{1}{\sqrt{f} H}
\left| \half  \partial_Hf \vpdl  \right| \, .
\ee

We close this section with some useful formulae for connecting the fields with the geometric quantities above
\bea
\dot{\sigma}^2&=&\gamma_{ab} \dot{\phi}^a \dot{\phi}^b, \el
T^a&=&\frac{\dot{\phi}^a}{\dot{\sigma}} \bskip \gamma_{ab}T^aT^b=1, \el
N^a&=&\frac{D_tT^a}{\sqrt{\gamma_{ab} D_tT^a D_tT^b}}, \el
\gamma_{ab}T^aN^b&=&0 \bskip \gamma_{ab}N^aN^b=1, \\
\nabla_b X^a &=& \partial_b X^a + \Gamma^a_{bc} X^c, \el
D_t X^a &\equiv& \frac{D}{dt} X^a=\frac{d}{dt} X^a + \Gamma^a_{bc} X^b \dot{\phi}^c, \el
D_t X^a &=& \dot{\sigma} T^b \nabla_b X^a = \dot{\sigma} \nabla_\phi X^a, \el
D_t \dot{\phi}^a&=& \ddot{\phi}^a + \Gamma^a_{bc} \dot{\phi}^b \dot{\phi}^c, \el
\ddot{\phi}^a&+&3H\dot{\phi}^a+\Gamma^{a}_{bc}  \dot{\phi}^b \dot{\phi}^c+\gamma^{ab}\partial_bV=0,
\eea
where we note that $\dot{\sigma}$ here is $\dot{\phi}_0$ in \cite{Shiu:2011qw}
corresponding to the adiabatic or tangent motion in field space.

\subsection{Connection with the Geometric Approach of Shiu and Xu}

Recall from the text that we can expand the light field around the last moment of adiabaticity, which gives:
\bea
\vpl(t) &\simeq & \vpl(r) + \vpdl(r) (t-r) - \frac{1}{6} | \vpdddl(r) | (t-r)^3,\nonumber \\
\vpdl(t)& \simeq& \vpdl(r)-\half  | \vpdddl(r) | (t-r)^2,\nonumber \\
\vpddl(t) &\simeq& - | \vpdddl(r) | (t-r),\nonumber \\
\vpdddl(t) & \simeq &  - | \vpdddl(r) |\nonumber \\
\vpdl^2(t) & \simeq & \vpdl^2(r) - \left| \vpdl(r) \vpdddl(r) \right| (t-r)^2 \, .
\label{scalarserieslast}
\eea
We note that these equations are approximations for small $\Delta t$ and, in particular, the
term $\vpdddl$ is only constant up to ${\cal O}(t-r)$.
We also recall that the interval during which particle production takes place is given by
\be
(\Delta t)^4 \sim \frac{\Lambda^2}{\vpdl(r) \vpdddl(r)}\, ,
\ee
with $\Delta t$ the upper bound on the time period of non-adiabaticity,
i.e. $t-r \lesssim \Delta t$ if we are interested in physics during the violation of adiabaticity.

Given these basic formulae, and the fact that we are interested in at most a small turn $\dt$ in field space in the heavy field direction
(we neglect terms of order $\dt^2$ or smaller),
we find the following relationships between the light and heavy field and the parameters of \cite{Shiu:2011qw}:
\bea
e^a_{\zeta}&=&( \dt, f^{-1/2}), \\
e^a_s &=& (-1, f^{-1/2} \dt), \\
\dt &=& \frac{\vpdh}{\dot{\varphi}_0} \simeq f^{-1/2} \frac{\vpdh}{\vpdl}, \\
\cD_te^H_{\zeta}&=& \dot{\dt} - \half \partial_H  f f^{-1/2} \vpdl, \\
\cD_te^L_{\zeta}&=&\half \partial_H \ln f  \vpdl \dt \, .
\ea
The turn rate $\dot\theta/H$ of \cite{Shiu:2011qw} is related to our parameters via:
\bea
\dot{\theta}&=& -\dot{\dt} + \half \partial_H f f^{-1/2} \vpdl  \, ,
\ea
and is therefore non-vanishing even in the case of $\dt=0$.
Derivatives of the potential in the notation of \cite{Shiu:2011qw} are related to ours as follows:
\bea
V_\zeta&=& f^{-1/2} V_L + V_H \dt, \\
V_{\zeta \zeta}&=& f^{-1} V_{LL}, \\
V_{s}&=& -f^{-1/2} V_L \dt - V_H,\\
V_{ss}&=& V_{HH} \, ,
\ea
where we have assumed a separable potential for simplicity.
Finally, the remaining relevant terms in the perturbations equations of motion of
\cite{Shiu:2011qw} are mapped to ours by using:
\bea
M_{\sigma \sigma}&=&V_{ss} + \epsilon H^2 m_p^2 R, \\
&=& V_{HH} + \epsilon H^2 m_p^2 R, \\
m_\sigma^2&=&V_{HH} - \half \partial^2_Hf \vpdl^2, \\
\frac{ \dot{z} }{z} &=& \half \partial_H f f^{-1/2} \vpdl \dt + \frac{\vpddl}{\vpdl}, \\
\frac{\ddot{z}}{z} &=& -V_{\zeta \zeta} + \dot{\theta}^2
=-f^{-1} V_{LL} + \left( -\dot{\dt} + \half \partial_H f f^{-1/2} \vpdl  \right)^2 \, .
\eea


\begin{thebibliography}{71}
\expandafter\ifx\csname natexlab\endcsname\relax\def\natexlab#1{#1}\fi
\expandafter\ifx\csname bibnamefont\endcsname\relax
  \def\bibnamefont#1{#1}\fi
\expandafter\ifx\csname bibfnamefont\endcsname\relax
  \def\bibfnamefont#1{#1}\fi
\expandafter\ifx\csname citenamefont\endcsname\relax
  \def\citenamefont#1{#1}\fi
\expandafter\ifx\csname url\endcsname\relax
  \def\url#1{\texttt{#1}}\fi
\expandafter\ifx\csname urlprefix\endcsname\relax\def\urlprefix{URL }\fi
\providecommand{\bibinfo}[2]{#2}
\providecommand{\eprint}[2][]{\url{#2}}

\bibitem[{\citenamefont{Kaplan}(2005)}]{Kaplan:2005es}
\bibinfo{author}{\bibfnamefont{D.~B.} \bibnamefont{Kaplan}}
  (\bibinfo{year}{2005}), \eprint{nucl-th/0510023}.

\bibitem[{\citenamefont{Burgess}(2007)}]{Burgess:2007pt}
\bibinfo{author}{\bibfnamefont{C.}~\bibnamefont{Burgess}},
  \bibinfo{journal}{Ann.Rev.Nucl.Part.Sci.} \textbf{\bibinfo{volume}{57}},
  \bibinfo{pages}{329} (\bibinfo{year}{2007}), \eprint{hep-th/0701053}.

\bibitem[{\citenamefont{Brandenberger}(2008)}]{Brandenberger:2007qi}
\bibinfo{author}{\bibfnamefont{R.~H.} \bibnamefont{Brandenberger}},
  \bibinfo{journal}{Lect.Notes Phys.} \textbf{\bibinfo{volume}{738}},
  \bibinfo{pages}{393} (\bibinfo{year}{2008}), \eprint{hep-th/0701111}.

\bibitem[{\citenamefont{Kaloper
  et~al.}(2002{\natexlab{a}})\citenamefont{Kaloper, Kleban, Lawrence, Shenker,
  and Susskind}}]{Kaloper:2002cs}
\bibinfo{author}{\bibfnamefont{N.}~\bibnamefont{Kaloper}},
  \bibinfo{author}{\bibfnamefont{M.}~\bibnamefont{Kleban}},
  \bibinfo{author}{\bibfnamefont{A.}~\bibnamefont{Lawrence}},
  \bibinfo{author}{\bibfnamefont{S.}~\bibnamefont{Shenker}}, \bibnamefont{and}
  \bibinfo{author}{\bibfnamefont{L.}~\bibnamefont{Susskind}},
  \bibinfo{journal}{JHEP} \textbf{\bibinfo{volume}{0211}}, \bibinfo{pages}{037}
  (\bibinfo{year}{2002}{\natexlab{a}}), \eprint{hep-th/0209231}.

\bibitem[{\citenamefont{Kaloper
  et~al.}(2002{\natexlab{b}})\citenamefont{Kaloper, Kleban, Lawrence, and
  Shenker}}]{Kaloper:2002uj}
\bibinfo{author}{\bibfnamefont{N.}~\bibnamefont{Kaloper}},
  \bibinfo{author}{\bibfnamefont{M.}~\bibnamefont{Kleban}},
  \bibinfo{author}{\bibfnamefont{A.~E.} \bibnamefont{Lawrence}},
  \bibnamefont{and} \bibinfo{author}{\bibfnamefont{S.}~\bibnamefont{Shenker}},
  \bibinfo{journal}{Phys.Rev.} \textbf{\bibinfo{volume}{D66}},
  \bibinfo{pages}{123510} (\bibinfo{year}{2002}{\natexlab{b}}),
  \eprint{hep-th/0201158}.

\bibitem[{\citenamefont{Weinberg}(2008)}]{Weinberg:2008hq}
\bibinfo{author}{\bibfnamefont{S.}~\bibnamefont{Weinberg}},
  \bibinfo{journal}{Phys. Rev.} \textbf{\bibinfo{volume}{D77}},
  \bibinfo{pages}{123541} (\bibinfo{year}{2008}), \eprint{0804.4291}.

\bibitem[{\citenamefont{Creminelli et~al.}(2009)\citenamefont{Creminelli,
  D'Amico, Norena, and Vernizzi}}]{Creminelli:2008wc}
\bibinfo{author}{\bibfnamefont{P.}~\bibnamefont{Creminelli}},
  \bibinfo{author}{\bibfnamefont{G.}~\bibnamefont{D'Amico}},
  \bibinfo{author}{\bibfnamefont{J.}~\bibnamefont{Norena}}, \bibnamefont{and}
  \bibinfo{author}{\bibfnamefont{F.}~\bibnamefont{Vernizzi}},
  \bibinfo{journal}{JCAP} \textbf{\bibinfo{volume}{0902}}, \bibinfo{pages}{018}
  (\bibinfo{year}{2009}), \eprint{0811.0827}.

\bibitem[{\citenamefont{Park et~al.}(2010)\citenamefont{Park, Zurek, and
  Watson}}]{Park:2010cw}
\bibinfo{author}{\bibfnamefont{M.}~\bibnamefont{Park}},
  \bibinfo{author}{\bibfnamefont{K.~M.} \bibnamefont{Zurek}}, \bibnamefont{and}
  \bibinfo{author}{\bibfnamefont{S.}~\bibnamefont{Watson}},
  \bibinfo{journal}{Phys.Rev.} \textbf{\bibinfo{volume}{D81}},
  \bibinfo{pages}{124008} (\bibinfo{year}{2010}), \eprint{1003.1722}.

\bibitem[{\citenamefont{Cheung et~al.}(2008)\citenamefont{Cheung, Creminelli,
  Fitzpatrick, Kaplan, and Senatore}}]{Cheung:2007st}
\bibinfo{author}{\bibfnamefont{C.}~\bibnamefont{Cheung}},
  \bibinfo{author}{\bibfnamefont{P.}~\bibnamefont{Creminelli}},
  \bibinfo{author}{\bibfnamefont{A.}~\bibnamefont{Fitzpatrick}},
  \bibinfo{author}{\bibfnamefont{J.}~\bibnamefont{Kaplan}}, \bibnamefont{and}
  \bibinfo{author}{\bibfnamefont{L.}~\bibnamefont{Senatore}},
  \bibinfo{journal}{JHEP} \textbf{\bibinfo{volume}{0803}}, \bibinfo{pages}{014}
  (\bibinfo{year}{2008}), \eprint{0709.0293}.

\bibitem[{\citenamefont{Senatore and
  Zaldarriaga}(2010{\natexlab{a}})}]{Senatore:2010wk}
\bibinfo{author}{\bibfnamefont{L.}~\bibnamefont{Senatore}} \bibnamefont{and}
  \bibinfo{author}{\bibfnamefont{M.}~\bibnamefont{Zaldarriaga}}
  (\bibinfo{year}{2010}{\natexlab{a}}), \eprint{1009.2093}.

\bibitem[{\citenamefont{O'Connell and Holman}(2011)}]{O'Connell:2011ng}
\bibinfo{author}{\bibfnamefont{R.}~\bibnamefont{O'Connell}} \bibnamefont{and}
  \bibinfo{author}{\bibfnamefont{R.}~\bibnamefont{Holman}}
  (\bibinfo{year}{2011}), \eprint{1109.1562}.

\bibitem[{\citenamefont{Senatore and Zaldarriaga}(2011)}]{Senatore:2010jy}
\bibinfo{author}{\bibfnamefont{L.}~\bibnamefont{Senatore}} \bibnamefont{and}
  \bibinfo{author}{\bibfnamefont{M.}~\bibnamefont{Zaldarriaga}},
  \bibinfo{journal}{JCAP} \textbf{\bibinfo{volume}{1101}}, \bibinfo{pages}{003}
  (\bibinfo{year}{2011}), \eprint{1004.1201}.

\bibitem[{\citenamefont{Senatore and
  Zaldarriaga}(2010{\natexlab{b}})}]{Senatore:2009cf}
\bibinfo{author}{\bibfnamefont{L.}~\bibnamefont{Senatore}} \bibnamefont{and}
  \bibinfo{author}{\bibfnamefont{M.}~\bibnamefont{Zaldarriaga}},
  \bibinfo{journal}{JHEP} \textbf{\bibinfo{volume}{1012}}, \bibinfo{pages}{008}
  (\bibinfo{year}{2010}{\natexlab{b}}), \eprint{0912.2734}.

\bibitem[{\citenamefont{Creminelli et~al.}(2011)\citenamefont{Creminelli,
  D'Amico, Musso, Norena, and Trincherini}}]{Creminelli:2010qf}
\bibinfo{author}{\bibfnamefont{P.}~\bibnamefont{Creminelli}},
  \bibinfo{author}{\bibfnamefont{G.}~\bibnamefont{D'Amico}},
  \bibinfo{author}{\bibfnamefont{M.}~\bibnamefont{Musso}},
  \bibinfo{author}{\bibfnamefont{J.}~\bibnamefont{Norena}}, \bibnamefont{and}
  \bibinfo{author}{\bibfnamefont{E.}~\bibnamefont{Trincherini}},
  \bibinfo{journal}{JCAP} \textbf{\bibinfo{volume}{1102}}, \bibinfo{pages}{006}
  (\bibinfo{year}{2011}), \eprint{1011.3004}.

\bibitem[{\citenamefont{Baumann and
  Green}(2011{\natexlab{a}})}]{Baumann:2011su}
\bibinfo{author}{\bibfnamefont{D.}~\bibnamefont{Baumann}} \bibnamefont{and}
  \bibinfo{author}{\bibfnamefont{D.}~\bibnamefont{Green}}
  (\bibinfo{year}{2011}{\natexlab{a}}), \eprint{1102.5343}.

\bibitem[{\citenamefont{Kaloper and Kaplinghat}(2003)}]{Kaloper:2003nv}
\bibinfo{author}{\bibfnamefont{N.}~\bibnamefont{Kaloper}} \bibnamefont{and}
  \bibinfo{author}{\bibfnamefont{M.}~\bibnamefont{Kaplinghat}},
  \bibinfo{journal}{Phys.Rev.} \textbf{\bibinfo{volume}{D68}},
  \bibinfo{pages}{123522} (\bibinfo{year}{2003}), \eprint{hep-th/0307016}.

\bibitem[{\citenamefont{Tolley and Wyman}(2010)}]{Tolley:2009fg}
\bibinfo{author}{\bibfnamefont{A.~J.} \bibnamefont{Tolley}} \bibnamefont{and}
  \bibinfo{author}{\bibfnamefont{M.}~\bibnamefont{Wyman}},
  \bibinfo{journal}{Phys.Rev.} \textbf{\bibinfo{volume}{D81}},
  \bibinfo{pages}{043502} (\bibinfo{year}{2010}), \eprint{0910.1853}.

\bibitem[{\citenamefont{Cremonini et~al.}(2010)\citenamefont{Cremonini, Lalak,
  and Turzynski}}]{Cremonini:2010sv}
\bibinfo{author}{\bibfnamefont{S.}~\bibnamefont{Cremonini}},
  \bibinfo{author}{\bibfnamefont{Z.}~\bibnamefont{Lalak}}, \bibnamefont{and}
  \bibinfo{author}{\bibfnamefont{K.}~\bibnamefont{Turzynski}},
  \bibinfo{journal}{Phys. Rev.} \textbf{\bibinfo{volume}{D82}},
  \bibinfo{pages}{047301} (\bibinfo{year}{2010}), \eprint{1005.4347}.

\bibitem[{\citenamefont{Cremonini et~al.}(2011)\citenamefont{Cremonini, Lalak,
  and Turzynski}}]{Cremonini:2010ua}
\bibinfo{author}{\bibfnamefont{S.}~\bibnamefont{Cremonini}},
  \bibinfo{author}{\bibfnamefont{Z.}~\bibnamefont{Lalak}}, \bibnamefont{and}
  \bibinfo{author}{\bibfnamefont{K.}~\bibnamefont{Turzynski}},
  \bibinfo{journal}{JCAP} \textbf{\bibinfo{volume}{1103}}, \bibinfo{pages}{016}
  (\bibinfo{year}{2011}), \eprint{1010.3021}.

\bibitem[{\citenamefont{Achucarro et~al.}(2011)\citenamefont{Achucarro, Gong,
  Hardeman, Palma, and Patil}}]{Achucarro:2010da}
\bibinfo{author}{\bibfnamefont{A.}~\bibnamefont{Achucarro}},
  \bibinfo{author}{\bibfnamefont{J.-O.} \bibnamefont{Gong}},
  \bibinfo{author}{\bibfnamefont{S.}~\bibnamefont{Hardeman}},
  \bibinfo{author}{\bibfnamefont{G.~A.} \bibnamefont{Palma}}, \bibnamefont{and}
  \bibinfo{author}{\bibfnamefont{S.~P.} \bibnamefont{Patil}},
  \bibinfo{journal}{JCAP} \textbf{\bibinfo{volume}{1101}}, \bibinfo{pages}{030}
  (\bibinfo{year}{2011}), \eprint{1010.3693}.

\bibitem[{\citenamefont{Shiu and Xu}(2011)}]{Shiu:2011qw}
\bibinfo{author}{\bibfnamefont{G.}~\bibnamefont{Shiu}} \bibnamefont{and}
  \bibinfo{author}{\bibfnamefont{J.}~\bibnamefont{Xu}} (\bibinfo{year}{2011}),
  \eprint{1108.0981}.

\bibitem[{\citenamefont{Chen and Wang}(2010)}]{Chen:2009zp}
\bibinfo{author}{\bibfnamefont{X.}~\bibnamefont{Chen}} \bibnamefont{and}
  \bibinfo{author}{\bibfnamefont{Y.}~\bibnamefont{Wang}},
  \bibinfo{journal}{JCAP} \textbf{\bibinfo{volume}{1004}}, \bibinfo{pages}{027}
  (\bibinfo{year}{2010}), \eprint{0911.3380}.

\bibitem[{\citenamefont{Senatore et~al.}(2011)\citenamefont{Senatore,
  Silverstein, and Zaldarriaga}}]{Senatore:2011sp}
\bibinfo{author}{\bibfnamefont{L.}~\bibnamefont{Senatore}},
  \bibinfo{author}{\bibfnamefont{E.}~\bibnamefont{Silverstein}},
  \bibnamefont{and}
  \bibinfo{author}{\bibfnamefont{M.}~\bibnamefont{Zaldarriaga}}
  (\bibinfo{year}{2011}), \eprint{1109.0542}.

\bibitem[{\citenamefont{Green et~al.}(2009)\citenamefont{Green, Horn, Senatore,
  and Silverstein}}]{Green:2009ds}
\bibinfo{author}{\bibfnamefont{D.}~\bibnamefont{Green}},
  \bibinfo{author}{\bibfnamefont{B.}~\bibnamefont{Horn}},
  \bibinfo{author}{\bibfnamefont{L.}~\bibnamefont{Senatore}}, \bibnamefont{and}
  \bibinfo{author}{\bibfnamefont{E.}~\bibnamefont{Silverstein}},
  \bibinfo{journal}{Phys.Rev.} \textbf{\bibinfo{volume}{D80}},
  \bibinfo{pages}{063533} (\bibinfo{year}{2009}), \eprint{0902.1006}.

\bibitem[{\citenamefont{Barnaby and Shandera}(2012)}]{Barnaby:2011pe}
\bibinfo{author}{\bibfnamefont{N.}~\bibnamefont{Barnaby}} \bibnamefont{and}
  \bibinfo{author}{\bibfnamefont{S.}~\bibnamefont{Shandera}},
  \bibinfo{journal}{JCAP} \textbf{\bibinfo{volume}{1201}}, \bibinfo{pages}{034}
  (\bibinfo{year}{2012}), \eprint{1109.2985}.

\bibitem[{\citenamefont{Barnaby}(2010)}]{Barnaby:2010ke}
\bibinfo{author}{\bibfnamefont{N.}~\bibnamefont{Barnaby}},
  \bibinfo{journal}{Phys.Rev.} \textbf{\bibinfo{volume}{D82}},
  \bibinfo{pages}{106009} (\bibinfo{year}{2010}), \eprint{1006.4615}.

\bibitem[{\citenamefont{Romano and Sasaki}(2008)}]{Romano:2008rr}
\bibinfo{author}{\bibfnamefont{A.~E.} \bibnamefont{Romano}} \bibnamefont{and}
  \bibinfo{author}{\bibfnamefont{M.}~\bibnamefont{Sasaki}},
  \bibinfo{journal}{Phys.Rev.} \textbf{\bibinfo{volume}{D78}},
  \bibinfo{pages}{103522} (\bibinfo{year}{2008}), \eprint{0809.5142}.

\bibitem[{\citenamefont{Jackson and Schalm}(2011)}]{Jackson:2011qg}
\bibinfo{author}{\bibfnamefont{M.~G.} \bibnamefont{Jackson}} \bibnamefont{and}
  \bibinfo{author}{\bibfnamefont{K.}~\bibnamefont{Schalm}}
  (\bibinfo{year}{2011}), \eprint{1104.0887}.

\bibitem[{\citenamefont{Martin and Sriramkumar}(2012)}]{Martin:2011sn}
\bibinfo{author}{\bibfnamefont{J.}~\bibnamefont{Martin}} \bibnamefont{and}
  \bibinfo{author}{\bibfnamefont{L.}~\bibnamefont{Sriramkumar}},
  \bibinfo{journal}{JCAP} \textbf{\bibinfo{volume}{1201}}, \bibinfo{pages}{008}
  (\bibinfo{year}{2012}), \eprint{1109.5838}.

\bibitem[{\citenamefont{Cespedes
  et~al.}(2012{\natexlab{a}})\citenamefont{Cespedes, Atal, and
  Palma}}]{Cespedes:2012}
\bibinfo{author}{\bibfnamefont{S.}~\bibnamefont{Cespedes}},
  \bibinfo{author}{\bibfnamefont{V.}~\bibnamefont{Atal}}, \bibnamefont{and}
  \bibinfo{author}{\bibfnamefont{G.~A.} \bibnamefont{Palma}}
  (\bibinfo{year}{2012}{\natexlab{a}}), \bibinfo{note}{32 pages, 10 figures},
  \eprint{1201.4848}.

\bibitem[{\citenamefont{Cook and Sorbo}(2011)}]{Cook:2011hg}
\bibinfo{author}{\bibfnamefont{J.~L.} \bibnamefont{Cook}} \bibnamefont{and}
  \bibinfo{author}{\bibfnamefont{L.}~\bibnamefont{Sorbo}}
  (\bibinfo{year}{2011}), \eprint{1109.0022}.

\bibitem[{\citenamefont{Park and Sorbo}(2012)}]{Park:2012rh}
\bibinfo{author}{\bibfnamefont{M.}~\bibnamefont{Park}} \bibnamefont{and}
  \bibinfo{author}{\bibfnamefont{L.}~\bibnamefont{Sorbo}}
  (\bibinfo{year}{2012}), \eprint{1201.2903}.

\bibitem[{\citenamefont{Battefeld et~al.}(2011)\citenamefont{Battefeld,
  Battefeld, Byrnes, and Langlois}}]{Battefeld:2011yj}
\bibinfo{author}{\bibfnamefont{D.}~\bibnamefont{Battefeld}},
  \bibinfo{author}{\bibfnamefont{T.}~\bibnamefont{Battefeld}},
  \bibinfo{author}{\bibfnamefont{C.}~\bibnamefont{Byrnes}}, \bibnamefont{and}
  \bibinfo{author}{\bibfnamefont{D.}~\bibnamefont{Langlois}},
  \bibinfo{journal}{JCAP} \textbf{\bibinfo{volume}{1108}}, \bibinfo{pages}{025}
  (\bibinfo{year}{2011}), \eprint{1106.1891}.

\bibitem[{\citenamefont{Peterson and
  Tegmark}(2011{\natexlab{a}})}]{Peterson:2010np}
\bibinfo{author}{\bibfnamefont{C.~M.} \bibnamefont{Peterson}} \bibnamefont{and}
  \bibinfo{author}{\bibfnamefont{M.}~\bibnamefont{Tegmark}},
  \bibinfo{journal}{Phys. Rev.} \textbf{\bibinfo{volume}{D83}},
  \bibinfo{pages}{023522} (\bibinfo{year}{2011}{\natexlab{a}}),
  \eprint{1005.4056}.

\bibitem[{\citenamefont{Avgoustidis et~al.}(2012)\citenamefont{Avgoustidis,
  Cremonini, Davis, Ribeiro, Turzynski, and Watson}}]{Avgoustidis:2011em}
\bibinfo{author}{\bibfnamefont{A.}~\bibnamefont{Avgoustidis}},
  \bibinfo{author}{\bibfnamefont{S.}~\bibnamefont{Cremonini}},
  \bibinfo{author}{\bibfnamefont{A.-C.} \bibnamefont{Davis}},
  \bibinfo{author}{\bibfnamefont{R.~H.} \bibnamefont{Ribeiro}},
  \bibinfo{author}{\bibfnamefont{K.}~\bibnamefont{Turzynski}},
  \bibnamefont{and} \bibinfo{author}{\bibfnamefont{S.}~\bibnamefont{Watson}},
  \bibinfo{journal}{JCAP} \textbf{\bibinfo{volume}{2012}}, \bibinfo{pages}{038}
  (\bibinfo{year}{2012}), \eprint{1110.4081}.

\bibitem[{\citenamefont{Starobinsky}(1992)}]{Starobinsky:1992ts}
\bibinfo{author}{\bibfnamefont{A.~A.} \bibnamefont{Starobinsky}},
  \bibinfo{journal}{JETP Lett.} \textbf{\bibinfo{volume}{55}},
  \bibinfo{pages}{489} (\bibinfo{year}{1992}).

\bibitem[{\citenamefont{Gong}(2005)}]{Gong:2005jr}
\bibinfo{author}{\bibfnamefont{J.-O.} \bibnamefont{Gong}},
  \bibinfo{journal}{JCAP} \textbf{\bibinfo{volume}{0507}}, \bibinfo{pages}{015}
  (\bibinfo{year}{2005}), \eprint{astro-ph/0504383}.

\bibitem[{\citenamefont{Adams et~al.}(2001)\citenamefont{Adams, Cresswell, and
  Easther}}]{Adams:2001vc}
\bibinfo{author}{\bibfnamefont{J.~A.} \bibnamefont{Adams}},
  \bibinfo{author}{\bibfnamefont{B.}~\bibnamefont{Cresswell}},
  \bibnamefont{and} \bibinfo{author}{\bibfnamefont{R.}~\bibnamefont{Easther}},
  \bibinfo{journal}{Phys.Rev.} \textbf{\bibinfo{volume}{D64}},
  \bibinfo{pages}{123514} (\bibinfo{year}{2001}), \eprint{astro-ph/0102236}.

\bibitem[{\citenamefont{Chen et~al.}(2007)\citenamefont{Chen, Easther, and
  Lim}}]{Chen:2006xjb}
\bibinfo{author}{\bibfnamefont{X.}~\bibnamefont{Chen}},
  \bibinfo{author}{\bibfnamefont{R.}~\bibnamefont{Easther}}, \bibnamefont{and}
  \bibinfo{author}{\bibfnamefont{E.~A.} \bibnamefont{Lim}},
  \bibinfo{journal}{JCAP} \textbf{\bibinfo{volume}{0706}}, \bibinfo{pages}{023}
  (\bibinfo{year}{2007}), \eprint{astro-ph/0611645}.

\bibitem[{\citenamefont{Baumann and
  Green}(2011{\natexlab{b}})}]{Baumann:2011nm}
\bibinfo{author}{\bibfnamefont{D.}~\bibnamefont{Baumann}} \bibnamefont{and}
  \bibinfo{author}{\bibfnamefont{D.}~\bibnamefont{Green}}
  (\bibinfo{year}{2011}{\natexlab{b}}), \eprint{1109.0293}.

\bibitem[{\citenamefont{Baumann and
  Green}(2011{\natexlab{c}})}]{Baumann:2011nk}
\bibinfo{author}{\bibfnamefont{D.}~\bibnamefont{Baumann}} \bibnamefont{and}
  \bibinfo{author}{\bibfnamefont{D.}~\bibnamefont{Green}}
  (\bibinfo{year}{2011}{\natexlab{c}}), \eprint{1109.0292}.

\bibitem[{\citenamefont{Achucarro et~al.}(2012)\citenamefont{Achucarro, Gong,
  Hardeman, Palma, and Patil}}]{Achucarro:2012sm}
\bibinfo{author}{\bibfnamefont{A.}~\bibnamefont{Achucarro}},
  \bibinfo{author}{\bibfnamefont{J.-O.} \bibnamefont{Gong}},
  \bibinfo{author}{\bibfnamefont{S.}~\bibnamefont{Hardeman}},
  \bibinfo{author}{\bibfnamefont{G.~A.} \bibnamefont{Palma}}, \bibnamefont{and}
  \bibinfo{author}{\bibfnamefont{S.~P.} \bibnamefont{Patil}}
  (\bibinfo{year}{2012}), \eprint{1201.6342}.

\bibitem[{\citenamefont{Cespedes
  et~al.}(2012{\natexlab{b}})\citenamefont{Cespedes, Atal, and
  Palma}}]{Cespedes:2012hu}
\bibinfo{author}{\bibfnamefont{S.}~\bibnamefont{Cespedes}},
  \bibinfo{author}{\bibfnamefont{V.}~\bibnamefont{Atal}}, \bibnamefont{and}
  \bibinfo{author}{\bibfnamefont{G.~A.} \bibnamefont{Palma}}
  (\bibinfo{year}{2012}{\natexlab{b}}), \eprint{1201.4848}.

\bibitem[{\citenamefont{Jackson and Schalm}(2012)}]{Jackson:2012fu}
\bibinfo{author}{\bibfnamefont{M.~G.} \bibnamefont{Jackson}} \bibnamefont{and}
  \bibinfo{author}{\bibfnamefont{K.}~\bibnamefont{Schalm}}
  (\bibinfo{year}{2012}), \eprint{1202.0604}.

\bibitem[{\citenamefont{Armendariz-Picon
  et~al.}(1999)\citenamefont{Armendariz-Picon, Damour, and
  Mukhanov}}]{ArmendarizPicon:1999rj}
\bibinfo{author}{\bibfnamefont{C.}~\bibnamefont{Armendariz-Picon}},
  \bibinfo{author}{\bibfnamefont{T.}~\bibnamefont{Damour}}, \bibnamefont{and}
  \bibinfo{author}{\bibfnamefont{V.~F.} \bibnamefont{Mukhanov}},
  \bibinfo{journal}{Phys.Lett.} \textbf{\bibinfo{volume}{B458}},
  \bibinfo{pages}{209} (\bibinfo{year}{1999}), \eprint{hep-th/9904075}.

\bibitem[{\citenamefont{Dong et~al.}(2011)\citenamefont{Dong, Horn,
  Silverstein, and Westphal}}]{Dong:2010in}
\bibinfo{author}{\bibfnamefont{X.}~\bibnamefont{Dong}},
  \bibinfo{author}{\bibfnamefont{B.}~\bibnamefont{Horn}},
  \bibinfo{author}{\bibfnamefont{E.}~\bibnamefont{Silverstein}},
  \bibnamefont{and} \bibinfo{author}{\bibfnamefont{A.}~\bibnamefont{Westphal}},
  \bibinfo{journal}{Phys.Rev.} \textbf{\bibinfo{volume}{D84}},
  \bibinfo{pages}{026011} (\bibinfo{year}{2011}), \eprint{1011.4521}.

\bibitem[{\citenamefont{Burgess
  et~al.}(2003{\natexlab{a}})\citenamefont{Burgess, Cline, and
  Holman}}]{Burgess:2003zw}
\bibinfo{author}{\bibfnamefont{C.}~\bibnamefont{Burgess}},
  \bibinfo{author}{\bibfnamefont{J.~M.} \bibnamefont{Cline}}, \bibnamefont{and}
  \bibinfo{author}{\bibfnamefont{R.}~\bibnamefont{Holman}},
  \bibinfo{journal}{JCAP} \textbf{\bibinfo{volume}{0310}}, \bibinfo{pages}{004}
  (\bibinfo{year}{2003}{\natexlab{a}}), \eprint{hep-th/0306079}.

\bibitem[{\citenamefont{Greene et~al.}(2004)\citenamefont{Greene, Schalm,
  van~der Schaar, and Shiu}}]{Greene:2005aj}
\bibinfo{author}{\bibfnamefont{B.}~\bibnamefont{Greene}},
  \bibinfo{author}{\bibfnamefont{K.}~\bibnamefont{Schalm}},
  \bibinfo{author}{\bibfnamefont{J.~P.} \bibnamefont{van~der Schaar}},
  \bibnamefont{and} \bibinfo{author}{\bibfnamefont{G.}~\bibnamefont{Shiu}},
  \bibinfo{journal}{eConf} \textbf{\bibinfo{volume}{C041213}},
  \bibinfo{pages}{0001} (\bibinfo{year}{2004}), \eprint{astro-ph/0503458}.

\bibitem[{\citenamefont{Silverstein and Tong}(2004)}]{Silverstein:2003hf}
\bibinfo{author}{\bibfnamefont{E.}~\bibnamefont{Silverstein}} \bibnamefont{and}
  \bibinfo{author}{\bibfnamefont{D.}~\bibnamefont{Tong}},
  \bibinfo{journal}{Phys.Rev.} \textbf{\bibinfo{volume}{D70}},
  \bibinfo{pages}{103505} (\bibinfo{year}{2004}), \eprint{hep-th/0310221}.

\bibitem[{\citenamefont{Traschen and Brandenberger}(1990)}]{Traschen:1990sw}
\bibinfo{author}{\bibfnamefont{J.~H.} \bibnamefont{Traschen}} \bibnamefont{and}
  \bibinfo{author}{\bibfnamefont{R.~H.} \bibnamefont{Brandenberger}},
  \bibinfo{journal}{Phys.Rev.} \textbf{\bibinfo{volume}{D42}},
  \bibinfo{pages}{2491} (\bibinfo{year}{1990}).

\bibitem[{\citenamefont{Kofman et~al.}(1997)\citenamefont{Kofman, Linde, and
  Starobinsky}}]{Kofman:1997yn}
\bibinfo{author}{\bibfnamefont{L.}~\bibnamefont{Kofman}},
  \bibinfo{author}{\bibfnamefont{A.~D.} \bibnamefont{Linde}}, \bibnamefont{and}
  \bibinfo{author}{\bibfnamefont{A.~A.} \bibnamefont{Starobinsky}},
  \bibinfo{journal}{Phys.Rev.} \textbf{\bibinfo{volume}{D56}},
  \bibinfo{pages}{3258} (\bibinfo{year}{1997}), \eprint{hep-ph/9704452}.

\bibitem[{\citenamefont{Birrell and Davies}(1982)}]{Birrell:1982ix}
\bibinfo{author}{\bibfnamefont{N.}~\bibnamefont{Birrell}} \bibnamefont{and}
  \bibinfo{author}{\bibfnamefont{P.}~\bibnamefont{Davies}}
  (\bibinfo{year}{1982}).

\bibitem[{\citenamefont{Burgess
  et~al.}(2003{\natexlab{b}})\citenamefont{Burgess, Cline, Lemieux, and
  Holman}}]{Burgess:2002ub}
\bibinfo{author}{\bibfnamefont{C.}~\bibnamefont{Burgess}},
  \bibinfo{author}{\bibfnamefont{J.~M.} \bibnamefont{Cline}},
  \bibinfo{author}{\bibfnamefont{F.}~\bibnamefont{Lemieux}}, \bibnamefont{and}
  \bibinfo{author}{\bibfnamefont{R.}~\bibnamefont{Holman}},
  \bibinfo{journal}{JHEP} \textbf{\bibinfo{volume}{0302}}, \bibinfo{pages}{048}
  (\bibinfo{year}{2003}{\natexlab{b}}), \eprint{hep-th/0210233}.

\bibitem[{\citenamefont{Barnaby et~al.}(2009)\citenamefont{Barnaby, Huang,
  Kofman, and Pogosyan}}]{Barnaby:2009mc}
\bibinfo{author}{\bibfnamefont{N.}~\bibnamefont{Barnaby}},
  \bibinfo{author}{\bibfnamefont{Z.}~\bibnamefont{Huang}},
  \bibinfo{author}{\bibfnamefont{L.}~\bibnamefont{Kofman}}, \bibnamefont{and}
  \bibinfo{author}{\bibfnamefont{D.}~\bibnamefont{Pogosyan}},
  \bibinfo{journal}{Phys.Rev.} \textbf{\bibinfo{volume}{D80}},
  \bibinfo{pages}{043501} (\bibinfo{year}{2009}), \eprint{0902.0615}.

\bibitem[{\citenamefont{Chung et~al.}(2011)\citenamefont{Chung, Everett, Yoo,
  and Zhou}}]{Chung:2011ck}
\bibinfo{author}{\bibfnamefont{D.~J.} \bibnamefont{Chung}},
  \bibinfo{author}{\bibfnamefont{L.~L.} \bibnamefont{Everett}},
  \bibinfo{author}{\bibfnamefont{H.}~\bibnamefont{Yoo}}, \bibnamefont{and}
  \bibinfo{author}{\bibfnamefont{P.}~\bibnamefont{Zhou}}
  (\bibinfo{year}{2011}), \eprint{1109.2524}.

\bibitem[{\citenamefont{Felder et~al.}(1999{\natexlab{a}})\citenamefont{Felder,
  Kofman, and Linde}}]{Felder:1998vq}
\bibinfo{author}{\bibfnamefont{G.~N.} \bibnamefont{Felder}},
  \bibinfo{author}{\bibfnamefont{L.}~\bibnamefont{Kofman}}, \bibnamefont{and}
  \bibinfo{author}{\bibfnamefont{A.~D.} \bibnamefont{Linde}},
  \bibinfo{journal}{Phys.Rev.} \textbf{\bibinfo{volume}{D59}},
  \bibinfo{pages}{123523} (\bibinfo{year}{1999}{\natexlab{a}}),
  \eprint{hep-ph/9812289}.

\bibitem[{\citenamefont{Felder et~al.}(1999{\natexlab{b}})\citenamefont{Felder,
  Kofman, and Linde}}]{Felder:1999pv}
\bibinfo{author}{\bibfnamefont{G.~N.} \bibnamefont{Felder}},
  \bibinfo{author}{\bibfnamefont{L.}~\bibnamefont{Kofman}}, \bibnamefont{and}
  \bibinfo{author}{\bibfnamefont{A.~D.} \bibnamefont{Linde}},
  \bibinfo{journal}{Phys.Rev.} \textbf{\bibinfo{volume}{D60}},
  \bibinfo{pages}{103505} (\bibinfo{year}{1999}{\natexlab{b}}),
  \eprint{hep-ph/9903350}.

\bibitem[{\citenamefont{Kofman et~al.}(2004)\citenamefont{Kofman, Linde, Liu,
  Maloney, McAllister et~al.}}]{Kofman:2004yc}
\bibinfo{author}{\bibfnamefont{L.}~\bibnamefont{Kofman}},
  \bibinfo{author}{\bibfnamefont{A.~D.} \bibnamefont{Linde}},
  \bibinfo{author}{\bibfnamefont{X.}~\bibnamefont{Liu}},
  \bibinfo{author}{\bibfnamefont{A.}~\bibnamefont{Maloney}},
  \bibinfo{author}{\bibfnamefont{L.}~\bibnamefont{McAllister}},
  \bibnamefont{et~al.}, \bibinfo{journal}{JHEP}
  \textbf{\bibinfo{volume}{0405}}, \bibinfo{pages}{030} (\bibinfo{year}{2004}),
  \eprint{hep-th/0403001}.

\bibitem[{\citenamefont{Watson}(2004)}]{Watson:2004aq}
\bibinfo{author}{\bibfnamefont{S.}~\bibnamefont{Watson}},
  \bibinfo{journal}{Phys.Rev.} \textbf{\bibinfo{volume}{D70}},
  \bibinfo{pages}{066005} (\bibinfo{year}{2004}), \eprint{hep-th/0404177}.

\bibitem[{\citenamefont{Cremonini and Watson}(2006)}]{Cremonini:2006sx}
\bibinfo{author}{\bibfnamefont{S.}~\bibnamefont{Cremonini}} \bibnamefont{and}
  \bibinfo{author}{\bibfnamefont{S.}~\bibnamefont{Watson}},
  \bibinfo{journal}{Phys.Rev.} \textbf{\bibinfo{volume}{D73}},
  \bibinfo{pages}{086007} (\bibinfo{year}{2006}), \eprint{hep-th/0601082}.

\bibitem[{\citenamefont{Greene et~al.}(2007)\citenamefont{Greene, Judes, Levin,
  Watson, and Weltman}}]{Greene:2007sa}
\bibinfo{author}{\bibfnamefont{B.}~\bibnamefont{Greene}},
  \bibinfo{author}{\bibfnamefont{S.}~\bibnamefont{Judes}},
  \bibinfo{author}{\bibfnamefont{J.}~\bibnamefont{Levin}},
  \bibinfo{author}{\bibfnamefont{S.}~\bibnamefont{Watson}}, \bibnamefont{and}
  \bibinfo{author}{\bibfnamefont{A.}~\bibnamefont{Weltman}},
  \bibinfo{journal}{JHEP} \textbf{\bibinfo{volume}{0707}}, \bibinfo{pages}{060}
  (\bibinfo{year}{2007}), \eprint{hep-th/0702220}.

\bibitem[{\citenamefont{Chung}(2003)}]{Chung:1998bt}
\bibinfo{author}{\bibfnamefont{D.~J.} \bibnamefont{Chung}},
  \bibinfo{journal}{Phys.Rev.} \textbf{\bibinfo{volume}{D67}},
  \bibinfo{pages}{083514} (\bibinfo{year}{2003}), \eprint{hep-ph/9809489}.

\bibitem[{\citenamefont{Lawrence and Martinec}(1996)}]{Lawrence:1995ct}
\bibinfo{author}{\bibfnamefont{A.~E.} \bibnamefont{Lawrence}} \bibnamefont{and}
  \bibinfo{author}{\bibfnamefont{E.~J.} \bibnamefont{Martinec}},
  \bibinfo{journal}{Class.Quant.Grav.} \textbf{\bibinfo{volume}{13}},
  \bibinfo{pages}{63} (\bibinfo{year}{1996}), \eprint{hep-th/9509149}.

\bibitem[{\citenamefont{Gubser}(2004)}]{Gubser:2003vk}
\bibinfo{author}{\bibfnamefont{S.~S.} \bibnamefont{Gubser}},
  \bibinfo{journal}{Phys.Rev.} \textbf{\bibinfo{volume}{D69}},
  \bibinfo{pages}{123507} (\bibinfo{year}{2004}), \eprint{hep-th/0305099}.

\bibitem[{\citenamefont{Lalak et~al.}(2007)\citenamefont{Lalak, Langlois,
  Pokorski, and Turzynski}}]{Lalak:2007vi}
\bibinfo{author}{\bibfnamefont{Z.}~\bibnamefont{Lalak}},
  \bibinfo{author}{\bibfnamefont{D.}~\bibnamefont{Langlois}},
  \bibinfo{author}{\bibfnamefont{S.}~\bibnamefont{Pokorski}}, \bibnamefont{and}
  \bibinfo{author}{\bibfnamefont{K.}~\bibnamefont{Turzynski}},
  \bibinfo{journal}{JCAP} \textbf{\bibinfo{volume}{0707}}, \bibinfo{pages}{014}
  (\bibinfo{year}{2007}), \eprint{0704.0212}.

\bibitem[{\citenamefont{Gordon et~al.}(2001)\citenamefont{Gordon, Wands,
  Bassett, and Maartens}}]{Gordon:2000hv}
\bibinfo{author}{\bibfnamefont{C.}~\bibnamefont{Gordon}},
  \bibinfo{author}{\bibfnamefont{D.}~\bibnamefont{Wands}},
  \bibinfo{author}{\bibfnamefont{B.~A.} \bibnamefont{Bassett}},
  \bibnamefont{and} \bibinfo{author}{\bibfnamefont{R.}~\bibnamefont{Maartens}},
  \bibinfo{journal}{Phys.Rev.} \textbf{\bibinfo{volume}{D63}},
  \bibinfo{pages}{023506} (\bibinfo{year}{2001}), \eprint{astro-ph/0009131}.

\bibitem[{\citenamefont{Di~Marco and Finelli}(2005)}]{DiMarco:2005nq}
\bibinfo{author}{\bibfnamefont{F.}~\bibnamefont{Di~Marco}} \bibnamefont{and}
  \bibinfo{author}{\bibfnamefont{F.}~\bibnamefont{Finelli}},
  \bibinfo{journal}{Phys.Rev.} \textbf{\bibinfo{volume}{D71}},
  \bibinfo{pages}{123502} (\bibinfo{year}{2005}), \eprint{astro-ph/0505198}.

\bibitem[{\citenamefont{Adshead et~al.}(2012)\citenamefont{Adshead, Dvorkin,
  Hu, and Lim}}]{Adshead:2011jq}
\bibinfo{author}{\bibfnamefont{P.}~\bibnamefont{Adshead}},
  \bibinfo{author}{\bibfnamefont{C.}~\bibnamefont{Dvorkin}},
  \bibinfo{author}{\bibfnamefont{W.}~\bibnamefont{Hu}}, \bibnamefont{and}
  \bibinfo{author}{\bibfnamefont{E.~A.} \bibnamefont{Lim}},
  \bibinfo{journal}{Phys.Rev.} \textbf{\bibinfo{volume}{D85}},
  \bibinfo{pages}{023531} (\bibinfo{year}{2012}), \eprint{1110.3050}.

\bibitem[{\citenamefont{Leblond and Shandera}(2008)}]{Leblond:2008gg}
\bibinfo{author}{\bibfnamefont{L.}~\bibnamefont{Leblond}} \bibnamefont{and}
  \bibinfo{author}{\bibfnamefont{S.}~\bibnamefont{Shandera}},
  \bibinfo{journal}{JCAP} \textbf{\bibinfo{volume}{0808}}, \bibinfo{pages}{007}
  (\bibinfo{year}{2008}), \eprint{0802.2290}.

\bibitem[{\citenamefont{Groot~Nibbelink and van
  Tent}(2002)}]{GrootNibbelink:2001qt}
\bibinfo{author}{\bibfnamefont{S.}~\bibnamefont{Groot~Nibbelink}}
  \bibnamefont{and} \bibinfo{author}{\bibfnamefont{B.}~\bibnamefont{van Tent}},
  \bibinfo{journal}{Class.Quant.Grav.} \textbf{\bibinfo{volume}{19}},
  \bibinfo{pages}{613} (\bibinfo{year}{2002}), \eprint{hep-ph/0107272}.

\bibitem[{\citenamefont{Peterson and
  Tegmark}(2011{\natexlab{b}})}]{Peterson:2011yt}
\bibinfo{author}{\bibfnamefont{C.~M.} \bibnamefont{Peterson}} \bibnamefont{and}
  \bibinfo{author}{\bibfnamefont{M.}~\bibnamefont{Tegmark}}
  (\bibinfo{year}{2011}{\natexlab{b}}), \eprint{1111.0927}.

\end{thebibliography}
\end{document}